\documentclass{JHEP3}
\usepackage{amsmath}
\usepackage{amssymb}
\usepackage{graphicx}
\numberwithin{equation}{section}

\usepackage{epsfig}

\renewcommand{\[}{\left[}
\renewcommand{\]}{\right]}
\renewcommand{\(}{\left(}
\renewcommand{\)}{\right)}

\newcommand{\mx}{x}

\newcommand{\px}{x^{\rm ph}}

\newcommand{\ph}{{\rm ph}}
\newcommand{\mir}{{\rm mir}}
\newcommand{\cut}{{\circlearrowright\hspace{-3.4mm}*\hspace{0.2mm}}}

\newcommand{\ff}{\scalebox{0.67}{\begin{picture}(5.21,0.0)\put(-1.5,1.2){$\oplus$}\end{picture}}}
\newcommand{\fF}{\scalebox{0.67}{\begin{picture}(5.21,0.0)\put(-1.5,1.2){$\otimes$}\end{picture}}}
\newcommand{\fm}{\scalebox{1.15}{\begin{picture}(2.8,0.0)\put(-1.2,-0.33){$\bullet$}\end{picture}}}
\newcommand{\fb}{\scalebox{0.5}{\begin{picture}(7.27,0.3)\put(-1.7,2.65){$\bigcirc$}\end{picture}}}
\newcommand{\fp}{\scalebox{0.6}[0.7]{\begin{picture}(7.24,0.2)\put(-1.5,0.5){$\bigtriangleup$}\end{picture}}}

\newcommand{\beq}{\begin{equation}}
\newcommand{\eeq}{\end{equation}}
\newcommand\beqa{\begin{eqnarray}}
\newcommand\eeqa{\end{eqnarray}}
\newcommand\bea{\begin{array}}
\newcommand\eea{\end{array}}

\newcommand\IM{{\rm Im}\,}
\newcommand\RE{{\rm Re}\,}

\def\XXint#1#2#3{{\setbox0=\hbox{$#1{#2#3}{\int}$}
\vcenter{\hbox{$#2#3$}}\kern-.5\wd0}}

\def\ccw{{\hspace{-2.5mm}\unitlength 0.1in
\begin{picture}(1.00,1.00)(8.67,-11.50)
\special{pn 8}%
\special{pa 1120 1120}%
\special{pa 1100 1100}%
\special{pa 1080 1120}%
\special{fp}%
\end{picture}%
\hspace{-0mm}}}

\def\cw{{\hspace{-2.5mm}\unitlength 0.1in
\begin{picture}(1.00,1.00)(8.67,-11.50)
\special{pn 8}%
\special{pa 1120 1100}%
\special{pa 1100 1120}%
\special{pa 1080 1100}%
\special{fp}%
\end{picture}%
\hspace{-0mm}}}

\newcommand{\nn}{\nonumber}

\newcommand{\COMMENT}[1]{}

\newcommand{\neqa}{\nonumber\end{eqnarray}}
\newcommand{\la}[1]{\label{#1}}

\newcommand{\eq}[1]{(\ref{#1})}

\def\tr{{\rm tr~}}
\newcommand{\Tr}{{\rm Tr}}

\renewcommand{\d}{\partial}

\newcommand{\<}{{\langle}}
\renewcommand{\>}{{\rangle}}

\newcommand{\re}{\relax{\rm I\kern-.18em R}}

\def\su2{{SU(2)}}

\def\<{\langle}
\def\>{\rangle}

\def\i2{\frac{i}{2}}

\title{Y-system and Quasi-Classical Strings}
\author{ Nikolay Gromov\\ DESY Theory, Hamburg, Germany \& II. Institut f\"{u}r Theoretische Physik Universit\"{a}t, Hamburg, Germany \&\\ St.Petersburg INP, St.Petersburg, Russia \\
E-mail: \email{nikgromov@gmail.com}}
\abstract{
Recently Kazakov, Vieira and the author
conjectured the $Y$-system set of equations describing the
planar spectrum of AdS/CFT.
In this paper we solve the $Y$-system equations
in the strong coupling scaling limit.
We show that the quasi-classical spectrum
of string moving inside $AdS_3\times S^1$
matches precisely with the prediction of the $Y$-system.
Thus the $Y$-system, unlike the
asymptotic Bethe ansatz, describes correctly the spectrum of one-loop string energies
including all exponential finite size corrections.
This gives a very non-trivial further support in favor of the conjecture.
}
\keywords{AdS/CFT, Integrability}
\preprint{}
\begin{document}



\section{Introduction}
Since the discovery of AdS/CFT  correspondence \cite{Maldacena:1997re,GKP}
there has been a significant progress in solving the maximally super-symmetric
Yang-Mills theory in four dimensions,
mainly due to the integrability found on both sides of the duality
\cite{Minahan:2002ve,Bena:2003wd,Kazakov:2004qf,Arutyunov:2004vx,Beisert:2005bm}.
The problem of finding the
anomalous dimensions of all local operators
in planar limit was thus brought very close to its complete solution.
Recently the anomalous dimension of
the simplest operator was computed numerically
for a wide range of 't Hooft coupling $\lambda$ \cite{Gromov:2009zb}.

The approach used there is based on the $Y$-system
for AdS/CFT conjectured in \cite{Gromov:2009tv}.
The main ingredients leading to the conjecture are
the asymptotic Bethe ansatz equations (ABA) \cite{Beisert:2006ez,Beisert:2005fw}, describing the spectrum
of the anomalous dimensions of local operators with
infinitely many constituent fields,
the concept of the ``mirror" double Wick rotated theory
invented in \cite{Ambjorn:2005wa} and then explored in details in \cite{Arutyunov:2007tc},
L\"uscher formula for the finite size corrections \cite{Bajnok:2008bm}
and our experience with relativistic theories, where similar equations
are constantly appearing
in the thermodynamic Bethe ansatz (TBA) approach
\cite{Zamolodchikov:1991et,Gromov:2008gj}.

This conjecture passes some nontrivial tests --
in \cite{Gromov:2009tv} the $4$-loop perturbative
result \cite{Fiamberti:2008sh} was reproduced.
More recently a comparison was also made at
$5$-loops in \cite{Fiamberti:2009jw}. In
\cite{Gromov:2009bc,Bombardelli:2009ns}
it was also shown to be consistent with the TBA approach
and finally its numerical solution \cite{Gromov:2009zb} indicates that the
strong coupling asymptotic agrees with the string prediction \cite{GKP}
for the simplest Konishi operator\footnote{The sub-sub-leading coefficient does not agree with
the quasi-classical string results of \cite{Roiban:2009aa}.
However, the Konishi state is one of the lowest states and the quasi-classical quantization may not be applicable for this case.}.
Apparently more tests of this conjecture are indispensable.
In this paper we compare the results of the quasi-classical string quantization
with the prediction of $Y$-system
thus providing the first analytical test of the conjecture to all orders in finite size
wrapping corrections. As we will see below this check involves numerous remarkable identities
and miraculous simplifications and probes most of the Y-system in great detail.
Therefore, the match we will observe leaves little room for doubt about the validity of
the Y-system, at least in the strong coupling limit.

The $Y$-system of \cite{Gromov:2009tv} is an infinite system of
simple functional equations\footnote
{In this paper we use a rescaled rapidities $z=\frac{u}{2g}$, where $g=\tfrac{\sqrt\lambda}{4\pi}$,
more convenient for our purposes.}
\beq\la{Ysys}
Y_{as}(z+\tfrac{i}{4g})Y_{as}(z-\tfrac{i}{4g})=
\frac{(1+Y_{a,s+1}(z))(1+Y_{a,s-1}(z))}{(1+1/Y_{a+1,s}(z))(1+1/Y_{a-1,s}(z))}\;,
\eeq
The indices $a,s$ belong to a T-shaped lattice (see Fig.\ref{fig1}).
One should replace $Y_{as}$ with indices outside the lattice
by $0$ or $\infty$ so that they disappear from the equations.
Note that the Y-system equations only involve the pair of indices $a,s$
and its immediate neighbors only, $a\pm1,s$ and $a,s\pm1$; in
this sense these equations are described as ``local".
In accordance with Fig.\ref{fig1} we also use the following
notations\footnote{In the original derivation of the Y-system equations the Y-functions
are associated with densities of several
bound states of the mirror theory \cite{Arutyunov:2009zu}.
The symbols used as subscripts in this new notation were introduced in
\cite{Gromov:2009bc} to indicate which type of bound-state they originate from.}
\begin{figure}[t]
\begin{center}
\includegraphics[width=60mm]{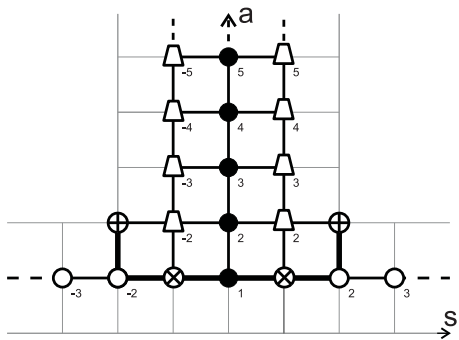}
\end{center}\la{fig1}
\caption{\textbf{T}-shaped ``fat hook". It defines the interactions between
$Y$-functions in the AdS/CFT $Y$-system equations.
}
\end{figure}
\beqa\la{nota}
\{Y_{\fm_a},Y_{\fp_a},Y_{\fb_s},Y_{\fF},Y_{\ff}\}
=\{Y_{a,0},
Y_{a,1},Y_{1,s},Y_{1,1},Y_{2,2}\}\;.
\eeqa

To have a unique solution this system should be
supplemented with an additional ``non-local" equation \cite{Gromov:2009bc}\footnote{We strongly
believe that Hirota equation associated with this $Y$-system
should automatically imply \eq{Ynl0}. From this point of view
Hirota equation seems to be more fundamental.
See also discussion in \cite{Gromov:2009tv}.}
\beq\la{Ynl0}
\log Y_{\fF}Y_{\ff}=
\sum_{a=1}^\infty\({\cal R}^{(0a)}-{\cal B}^{(0a)}\)*\log(1+Y_{\fm_a})+\log\frac{R^{(+)}B^{(-)}}{R^{(-)}B^{(+)}}\;,
\eeq
and particular boundary conditions at infinite $a$ and $s$. Here,
$\cal R$ and $\cal B$ are some kernels defined in Appendix A and $*$ stands for convolution.
This additional data encodes information about particular
operator/state.

When the solution is found the energy can be computed from
\beq
\la{Energy}
E=\sum_{j=1}^M\epsilon_1^{\ph}(z_j)+
\sum_{a=1}^\infty\int_{-\infty}^\infty
\frac{dz}{2\pi i}\frac{\d\epsilon_a^{\mir}(z)}{\d z}\log(1+Y^{\mir}_{\fm_a})\;,
\eeq
where $\epsilon_a(z)$ is a single magnon energy\footnote{Magnons are the fundamental spin
chain excitations around the BPS vacuum $\tr Z^L$. They are dual to the string
worldsheet excitations. Their dispersion relation and the scattering
matrix describing magnon scattering are known at any coupling and are properties
of the infinite volume theory.}
It is a simple multi-valued function and $\ph/\mir$ indicate its
branches (see Appendix A for notations).
Similarly $Y_{\fm_a}^\ph$  and $Y_{\fm_a}^\mir$
correspond to the different branches.
Finally $z_j$ are exact Bethe roots satisfying the exact Bethe equation
\beq\la{BAEex}
Y^{\ph}_{\fm_1}(z_i)=-1\;.
\eeq

In this paper we will focus on the $\frak{sl}(2)$
subsector of the correspondence. From the gauge theory point of view
this subsector is a closed subsector, with operators being composed from scalar fields
$\cal Z$ and covariant derivatives $\cal D$,
\beq
\Tr\({\cal D}^M {\cal Z}^L\)+\dots\; .
\eeq
The dots stand for all possible permutations of the derivatives with $\cal Z$ fields.
In the Y-system language,  $M$ is the number of roots $z_j$ whereas $L$ enters through large $a$ asymptotic of $Y_{\fm_a}$.
From the string theory side, this subsector describes strings moving in $AdS_3\times S^1$ contained in the larger $AdS_5\times S^5$ space-time. At infinite coupling the string motion is
classical while the first $1/\sqrt{\lambda}$ corrections are the quasi-classical one-loop effects which we study in this paper.
For this subsector the solution is $s\to -s $ symmetric $Y_{a,s}=Y_{a,-s}$,
which makes our consideration more transparent.

In the limit when the number of operators $L$
goes to infinity, $Y_{\fm_a}^\mir$ are exponentially suppressed
and the integral term in \eq{Energy} becomes irrelevant.
The remaining term is the sum of the individual energies of magnons.
In this limit the solution of $Y$-system is known for arbitrary coupling $g=\frac{\sqrt{\lambda}}{4\pi}$
and arbitrary state \cite{Gromov:2009tv}.
The finite size effects enter in two different ways: firstly,
the second term in \eq{Energy} becomes important and, secondly,
the Bethe roots $z_i$ move away from their asymptotic values
giving rise to the modification of the first term.

In this paper we solve the $Y$-system equations \eq{Ysys} in the strong coupling limit.
We construct explicitly the $Y$-functions
for large number of Bethe roots $M\sim g\gg 1$.
Then we compare the results with the quasi-classically quantized string
using methods developed in \cite{BF,Gromov:2007ky,Gromov:2007cd,Gromov:2007aq,Gromov:2008ec,Gromov:2008ie,Gromov:2009zz}
and show their perfect match.
We show that the $Y$-system resolves the know disagreement of asymptotic Bethe ansatz
with the semi-classically quantized strings \cite{SchaferNameki:2006gk}
due to finite-size effects.

\section{$Y$-system in the scaling limit}
The scaling limit is the strong coupling limit
$g\to\infty$ where the string can be described classically.
In this limit, the number of Bethe roots $M$
and the operator length $L$ go to infinity as $g$.
We also assume that the Bethe roots $z_i\sim 1$ and $|z_i|>1$.
This limit of Bethe equations was introduced
 in \cite{Sutherland:1995zz,Beisert:2003xu}.
The Bethe roots form continuous cuts in the complex plane $z$.
They become the branch-cuts on the classical algebraic curve which we describe in Sec.\ref{sec:ac}.

In the scaling limit we can neglect $i/(4g)$ shifts in the arguments in l.h.s.
of \eq{Ysys} which with $1/g^2$ precision
becomes a set of algebraic equations. This is the key simplification which will allow us to
completely solve the Y-system at strong coupling.
In the notations \eq{nota} we have three
infinite series of equations
\beqa
\la{ie1}Y_{\fb_s}^2&=&(1+Y_{\fb_{s+1}})(1+Y_{\fb_{s-1}})\;\;,\;\;s=3,4,\dots\;,\\
\la{ie2}Y_{\fp_a}^2&=&\frac{(1+Y_{\fm_a})}{(1+1/Y_{\fp_{a+1}})(1+1/Y_{\fp_{a-1}})}\;\;,\;\;a=3,4,\dots\;,\\
\la{ie3}Y_{\fm_a}^2&=&\frac{(1+Y_{\fp_a})^2}{(1+1/Y_{\fm_{a+1}})(1+1/Y_{\fm_{a-1}})}\;\;,\;\;a=2,3,\dots\;,
\eeqa
plus four equations
\beqa
\la{ne1}Y_{\fp_2}^2&=&\frac{(1+Y_{\ff})(1+Y_{\fm_2})}{(1+1/Y_{\fp_3})(1+1/Y_{\fF})}\;,\\
\la{ne2}Y_{\fb_2}^2&=&\frac{(1+Y_{\fb_3})(1+Y_{\fF})}{(1+1/Y_{\ff})}\;,\\
\la{ne3}Y_{\fF}^2&=&\frac{(1+Y_{\fb_2})(1+Y_{\fm_1})}{(1+1/Y_{\fp_2})}\;,\\
\la{ne4}Y_{\fm_1}^2&=&\frac{(1+Y_{\fF})^2}{(1+1/Y_{\fm_{2}})}\;.
\eeqa

From the asymptotic solution of \cite{Gromov:2009tv}, which we re-consider in the next section,
one expects that $Y$-functions have several branch cuts going from
$z=\pm 1+\frac{in}{4g}$, for some integers $n$, to infinity.
The approximation we use above is only accurate
far enough from the branch-cuts. Close to the branch-cuts even a small
shift in the argument could cause a large jump of $Y$'s.
This means we can safely use (\ref{ie1}-{\ref{ne4})
above and below the real axis and
on the interval $[-1,1]$ of the real axis, but not
on the whole real axis.
In this section we will mainly focus on the spectral parameters
with $-1<\RE z<1$ for which (\ref{ie1}-{\ref{ne4}) are valid.

Equation \eq{Ynl0} also simplifies in the scaling limit
\beq
\frac{R^{(+)}B^{(-)}}{R^{(-)}B^{(+)}}=
\prod_{j=1}^M\frac{x(z)-x_j^-}{x(z)-x_j^+}\frac{1/x(z)-x_j^+}{1/x(z)-x_j^-}
\simeq \frac{1}{f(z)\bar f(z)}\;,
\eeq
where
\beq
x(z)=z+i\sqrt{1-z^2}\;\;,\;\;x_j=x^{\ph}(z_j)=z_j+\sqrt{z_j-1}\sqrt{z_j+1}
\eeq
and $x_j^{\pm}=x^{\ph}(z_j\pm \tfrac{i}{4g})$
for the standard definition of $\sqrt{\dots}$ with branch cut along negative part of the
real axis. The functions $f(z)$ and $\bar f(z)$ are defined in terms of the resolvent
\beq\la{ResDef}
G(x)=\frac{1}{g}\sum_{j}^M\frac{1}{x-x_j}\frac{x_j^2}{x_j^2-1}\;,
\eeq
in the following way
\beq
f(z)=\exp\Big(-i G\big(x(z)\big)\Big)
\;\;,\;\;
\bar f(z)=\exp\Big(+i G\big(1/x(z)\big)\Big)\;. \la{fs}
\eeq
The kernels ${\cal R}^{(0a)}$ and ${\cal B}^{(0a)}$, defined in Appendix A
simplify dramatically in the strong coupling scaling limit.
For example ${\cal R}^{(0a)}(z,w)-{\cal B}^{(0a)}(z,w)\simeq\delta(z-w)$ 
so that equation \eq{Ynl0} becomes
\beq\la{Ynl}
F\equiv Y_{\fF}Y_{\ff}=
\frac{1}{f\bar f}\prod_{a=1}^\infty\log(1+Y_{\fm_a})
\;.
\eeq

One should keep in mind that the possible distribution of the Bethe roots is constrained by
\beq\la{zeromom0}
\sum_{j=1}^M\frac{1}{i}\log\frac{x_j^+}{x_j^-}\simeq \sum_{j=1}^M\frac{x_j}{g(x_j^2-1)}
=2\pi m+{\cal O}(1/g)\;\;,\;\;m\in{\mathbb Z}\;,
\eeq
which reflects the cyclicity symmetry of the single trace operators.

We see that in the scaling limit we have to solve an infinite set
of algebraic equations (\ref{ie1}-\ref{ne4},\ref{Ynl}).
In the next section we construct the asymptotic solution of these equations for $Y_{\fm_a}\ll 1$
and expand it in the scaling limit.

\subsection{Infinite length solution at strong coupling}\la{sec:nw}
In this section we will study the $Y$-system in the asymptotic limit $Y_{\fm_a}\ll 1$
and then expand it at strong coupling. We will see that in this scaling limit the asymptotic solutions of  \cite{Gromov:2009tv} can be recast in the following simple form:
\beq\la{Ynw}
Y_{\fb_s}(z)=\(s-A\)^2-1\;\;,\;\;
Y_{\fp_a}
=\frac{(T-1)^2 S T^{a-1}}{(S T^{a+1}-1)(S T^{a-1}-1)}\;,
\eeq
where
\beq\la{A0}
A=\frac{1}{\bar{f}-1}+\frac{f}{f-1}\;\;,\;\;
S=\frac{\bar f(f-1)^2}{f(\bar f-1)^2}\;\;,\;\;
T=\frac{f}{\bar f}\;.
\eeq
with (\ref{fs}). The middle node Y-functions are equally simple:
\beq
Y_{\fm_a}
=
 \Delta^a\(\frac{f(\bar f-1)^2\bar f^a-(f-1)^2\bar f f^a}{f\bar f(\bar f-f)}\)^2\;, \la{Ynw2}
\eeq
where
\beq\la{DELTA}
\Delta=\exp\(-\frac{\frac{L}{2g}+2\pi m z}{\sqrt{1-z^2}}\)\;,
\eeq
governs their exponential suppression for large $L/g$. We now derive these equations starting from the asymptotic solution constructed in \cite{Gromov:2009tv} in terms of the eigenvalues of
$SU(2|2)$ transfer matrices for representations with rectangular Young tableaux $T_{as}$. We have
\beq
1+1/Y_{\fb_s}(z)=\frac{T_{1,s}(z+\tfrac{i}{4g})T_{1,s}(z-\tfrac{i}{4g})}{T_{1,s-1}(z)T_{0,s+1}(z)}
\;\;,\;\;
1+Y_{\fp_a}(z)=\frac{T_{a,1}(z+\tfrac{i}{4g})T_{a,1}(z-\tfrac{i}{4g})}{T_{a-1,1}(z)T_{a+1,1}(z)}\;.
\eeq
The middle node functions $Y_{\fm_a}(z)$ are suppressed but can be also expressed
in terms of the transfer matrices using
\beq\label{eq:AssY}
Y_{\fm_a}(z)= \(T_{a,1}\)^2\left(\frac{x(z-\tfrac{ia}{4g})}{x(z+\tfrac{ia}{4g})}\right)^{L}
\prod_{b=-\frac{a-1}{2}}^{\frac{a-1}{2}}\prod_{j=1}^M
\frac{ B^{(+)}(z+i\tfrac{2b+1}{4g})R^{(-)}(z+i\tfrac{2b-1}{4g})}{B^{(-)}(z+i\tfrac{2b-1}{4g})R^{(+)}(z+i\tfrac{2b+1}{4g})}
\sigma^2(z+\tfrac{ib}{2g},x_j)
\;,
\eeq
where $B^{(\pm)},\;R^{(\pm)}$ are defined in the Appendix A
and $\sigma(z,x_j)$ for $\IM z>0$ coincides
with the Beisert-Eden-Staudacher dressing phase
\cite{Beisert:2006ez} and is understood as an analytical
continuation otherwise.
The $SU(2|2)$ transfer matrices for symmetric ($T_{1,s}$) and antisymmetric ($T_{a,1}$) representations can be found from the expansion of the generating
functional \cite{Tsuboi:1997iq,Kazakov:2007fy} (see \cite{Arutyunov:2009ce} for some mathematical details). For the $\frak{sl}(2)$ subsector it reads
\beq
{\cal W}=
\left[1-\frac{B^{(+)}(z+\tfrac{i}{4g})R^{(+)}(z-\tfrac{i}{4g}) }{B^{(-)}(z+\tfrac{i}{4g})R^{(-)}(z-\tfrac{i}{4g}) }
\hat D\right]\left[1-(-1)^m\frac{R^{(+)}(z-\tfrac{i}{4g}) }{ R^{(-)}(z-\tfrac{i}{4g}) } \hat D\right]^{-2}\left[1- \hat D\right]\;,
\la{WW}
\eeq
where $\hat D=e^{-i\frac{\partial_z}{2g}}$ is a shift operator. One should expand \eq{WW}
in powers of $\hat D$ and commute them to the right
\beq
{\cal W}=\sum_{s=0}^\infty
T_{1s}\(z-i\tfrac{s-1}{4g}\) \hat D^s \,\, , \,\,
{\cal W}^{-1}=\sum_{a=0}^\infty (-1)^a
T_{a1}\(z-i\tfrac{a-1}{4g}\)\hat D^a\;.
\eeq
We want to construct $Y_{as}$ using this construction
in the scaling limit.
At strong coupling the shift operator $\hat D$
serves just as a formal expansion parameter $D$, since
the shifts it creates are suppressed. In the notations of the previous section $\cal W$
becomes
\beq
{\cal W}=
\frac{\(1-D\bar f /f\)\(1-D\)}
{\left(1-D/f\right)^{2}}\;.
\eeq
Now it is very easy to find the general expression for $T_{1s}$ and $T_{a1}$
\beq\la{Tleading}
T_{1s}=\frac{(s-1)f\bar f-s(f+\bar f)+(s+1)}{f^s}\;\;,\;\;
T_{a1}=(-1)^a\frac{f(\bar f-1)^2\bar f^a-\bar f(f-1)^2 f^a}{f\bar f(\bar f-f)f^{a}}\;,
\eeq
which implies the above mentioned expressions (\ref{Ynw}) for the Y-functions.
In the expression for the middle node $Y$'s \eq{eq:AssY} it is enough to use the leading strong coupling
expression for the dressing phase -- the Arutyunov--Frolov--Staudacher (AFS) phase \cite{Arutyunov:2004vx}
\beq\la{AFS}
\sigma_a^2(z,x_j)\simeq
\(\frac{1-1/(x^-(z)x_j^+)}{1-1/(x^+(z)x_j^-)}\)^{2}
\(\frac{x^-(z)x_j^--1}{x^+(z)x_j^--1}\frac{x^+(z)x_j^+-1}{x^-(z)x_j^+-1}\)^{4ig(z_j-z)}\;,
\eeq
where $x^\pm=x(z\pm \tfrac{i}{4g})$. Expanding this expression at strong coupling for $-1<z<1$
we obtain $Y_{\fm_a} \simeq \(f^{2}\Delta\)^a\!T_{a1}^2$, or (\ref{Ynw2}).

We end this section with an important comment which will be later used in Sec.\ref{ExectBAE}.
We notice that for $|z|>1$ the second factor in \eq{eq:AssY} is exponentially small even for finite $L/g$
(notice that $x(z)$ denotes the mirror branch $x(z)=z+i\sqrt{1-z^2}$)
\beq\la{Ymsup}
\left(\frac{x(z-\tfrac{ia}{4g})}{x(z+\tfrac{ia}{4g})}\right)^{L}\simeq \frac{1}{x(z)^{aL}}\;,
\eeq
and thus the asymptotic solution is accurate for $|z|>1$ even if for $-1<z<1$ it is significantly modified by the finite size effects. Let us stress  that the exponential suppression we are discussing is much stronger than the usual finite size exponential suppression at strong coupling. The latter is suppressed for large $L/g$ whereas (\ref{Ymsup}) is suppressed for large $L$ even if $L/g$ is finite and small.

In the next section we will analyze the large $a$ and $s$ limit of these asymptotic $Y$'s
and argue that the same asymptotics should be used even when the fine size effects are strong.

\subsection{Boundary conditions}\la{sec:bc}
In this section we propose the boundary conditions which should be used
to make the solution of the $Y$-system unique. For that we study
the asymptotic large $L$ solution considered in the previous section
at large $a$ or $s$ and argue that the exact solution should have
exactly the same behavior.

From \eq{Ynw} we see that $Y_{\fp_a}$
oscillates with $a$ because $T(z)=f(z)/\bar f(z)$  is a pure phase.
To have a well defined large $a$ limit we shift the argument by $-i0$
then $|T(z-i0)|>1$ and we get
\beq\la{bp}
\lim_{a\to\infty}\frac{\log Y_{\fp_a}(z-i0)}{a}=\log \frac{\bar f}{f}\;.
\eeq
Similarly
\beq\la{bm}
\lim_{a\to\infty}\frac{\log Y_{\fm_a}(z-i0)}{a}=\log \(\Delta f^2\)\;.
\eeq
Whereas $Y_{\fp_a}$ and $Y_{\fm_a}$
decrease exponentially with $a$, $Y_{\fb_s}$ behave as $s^2$.
The general solution of \eq{ie1} with polynomial asymptotics
is
\beq\la{bb}
Y_{\fb_s}=(s-A(z))^2-1
\eeq
for some $A(z)$. When $\Delta$ is small and $Y_{\fm_a}$ are suppressed
 $A(z)$ is given by its asymptotic value \eq{A0}, otherwise it is some unknown function. In the next section
 we find
its exact expression as a function of $\Delta$.

Conditions \eq{bp} , \eq{bm} and \eq{bb} could be seen to be consistent with
the TBA equations for excited states proposed in \cite{Gromov:2009bc}
(see also \cite{Bombardelli:2009ns})
$Y_{{\fp}_{a}}$ should satisfy\footnote{the $-i0$ is due to the prescription to
go below singularities in the convolutions \cite{Gromov:2009bc}.}
\beq\nn
\log Y_{{\fp}_{a}}(z-i0)\!=\!{\cal M}_{am}\!*\!\log(1+Y_{\fm_m})\!-\!K_{a-1,m-1}\!*\!\log(1+Y_{{\fp}_{m}})
\!-\!K_{a-1}\!\cut\!\log(1+Y_{\fF})
\!+\!\log\!\frac{R_a^{(+)}\!{B_{a-2}^{(+)}}}{R_a^{(-)}\!B_{a-2}^{(-)}}
\eeq
where $\cal M$ and $K$ are some kernels defined in Appendix A. At strong coupling
the last term
gives
\beq
\lim_{a\to\infty}\frac{1}{a}\log\frac{R_a^{(+)}{B_{a-2}^{(+)}}}{R_a^{(-)}B_{a-2}^{(-)}}\simeq \log \frac{\bar f}{f}\;,
\eeq
assuming that the other terms are not growing with $a$ linearly this leads precisely to \eq{bp}. Similarly, \eq{bm} and \eq{bb} could be justified
from the TBA equations
of \cite{Gromov:2009bc}.

\subsection{Y-system in $T$-hook}
In this section we solve (\ref{ie1}-\ref{ne4}) together with
\eq{Ynl}.
One can achieve a considerable simplification of this problem
by transforming (\ref{ie1}-\ref{ne4}) into the Hirota equation.
For that we rewrite $Y_{as}$ in terms of $T_{as}$
\beq\la{YasT}
Y_{a,s}=\frac{T_{a,s+1}T_{a,s-1}}{T_{a+1,s}T_{a-1,s}}\;.
\eeq
$T_{as}$ should satisfy the Hirota equations
\beq
T_{a,s}^2=T_{a+1,s}T_{a-1,s}+T_{a,s+1}T_{a,s-1}\la{hir}\;,
\eeq
from which all (\ref{ie1}-\ref{ne4}) follow.
The indices of these $T_{as}$ functions
belong to the $T$-shaped lattice (see Fig.2).
Another equivalent representation, which follows from \eq{hir} is
\beq\la{TtoY2}
1+Y_{a,s}=\frac{T_{a,s}^2}{T_{a+1,s}T_{a-1,s}}\;.
\eeq

It is important to notice that the choice of $T_{as}$
is not unique for given $Y_{as}$, there is a ``gauge" freedom
\beq\la{gauge}
T_{a,s}(z)\to g_1(z) [g_2(z)]^a T_{a,s}(z)\;,
\eeq
which leaves $Y$'s unchanged for two arbitrary functions $g_1(z)$ and $g_2(z)$.

Below we will see how the general solution of \eq{hir}
can be constructed for the infinite vertical strip,
which is the upper part of the $T$-hook.
Then we constrain it by the large $a$ asymptotic
\eq{bp} and \eq{bm}, and match with
$Y_{\fb_s}$ given by \eq{bb} (shown as dark gray circles on Fig.2). As a result all $Y$'s
are constructed explicitly in Sec.\ref{eq:Ysolved} for finite $L/g$.
\begin{figure}[t]\la{fig2}
\begin{center}
\includegraphics[width=100mm]{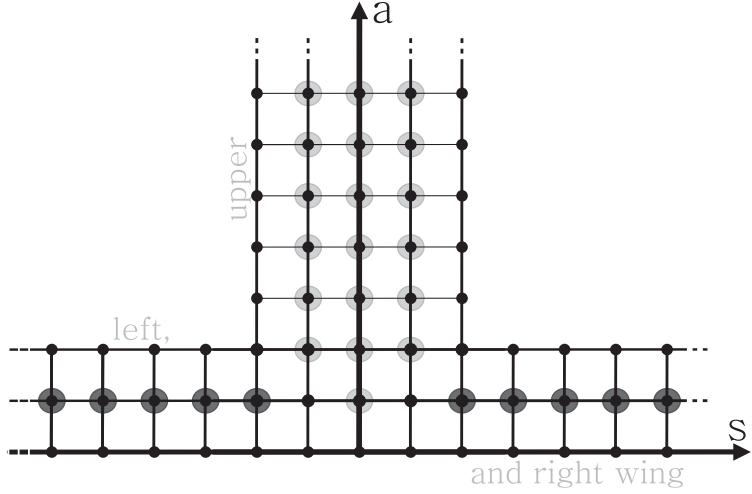}
\end{center}
\caption{T-shaped ``fat hook". Small black filled circles correspond to the finite $T_{as}$
function. Big circles show where solution for $Y$-functions for the strips is applicable.
}
\end{figure}

\subsubsection{Solution of Hirota equation in the vertical strip}
\la{sec:sol}

We notice that inside the vertical strip of Fig.2
the Hirota equation \eq{hir} coincides with the recurrent relation for the characters
of $SU(4)$ for the representations with rectangular Young tableaux.
It is well known that the characters are given
by Schur polynomials for Young tableau $(\lambda_1,\dots,\lambda_4)$:\footnote{We would like
to thank V.Kazakov for discussing this point.}
\beq\la{Shp}
s_{\lambda}(y)=\frac{\det (y_i^{\lambda_j+4-j})_{1\leq i,j\leq 4}}{\det (y_i^{4-j})_{1\leq i,j\leq 4}}\;.
\eeq
We only need symmetric solutions $T_{a,-s}=T_{a,s}$
which implies that $y_4=1/y_1$ and $y_3=1/y_2$. For rectangular representation $\lambda_{i\leq s+2}=0,\;\lambda_{i> s+2}=a$,
 from \eq{Shp} we get
$T_{a,2}=1$ and
\beqa
\la{Thr} T_{a,1}&=&\frac{y_1y_2}{(y_1-y_2)(y_1y_2-1)}\(\frac{y_1}{y_1^2-1}\(\frac{1}{y_1^{a-2}}-y_1^{a-2}\)
-\frac{y_2}{y_2^2-1}\(\frac{1}{y_2^{a-2}}-y_2^{a-2}\)
\).
\eeqa
This solution of \eq{hir} has only $2$ parameters because it is suitable for a finite semi-infinite strip boundary conditions, appropriate for characters. On the other hand, the most general $s\to-s$ symmetric solution should have
$4$ free parameters. It is clear that we can get one more by shifting $a$ in
\eq{Thr} by an arbitrary function. What is also true, although not as trivial, is that we can
shift independently $a$ in $y_1^a$ and in $y_2^a$ so that the most
general solution of \eq{hir} is
\beqa
\nn T_{a,2}&=&1\;,\\
\la{Tgen} T_{a,1}&=&\frac{iy_1y_2}{(y_1-y_2)(y_1y_2-1)}\(\frac{y_1}{y_1^2-1}\(\frac{1}{S_1y_1^{a}}+S_1y_1^{a}\)
-\frac{y_2}{y_2^2-1}\(\frac{1}{S_2y_2^{a}}+S_2y_2^{a}\)
\)\;,\\
\nn T_{a,0}&=&T_{a,1}^2-T_{a+1,1}T_{a-1,1}\;,
\eeqa
all the others $T$'s are given by $T_{a,-s}=T_{a,s}$.

To establish a relation with the previous section we reparameterize our solution in terms of
$4$ new parameters in the following way
\beq
y_1=\sqrt{\epsilon}\;\;,\;\;y_2=\frac{\sqrt{\epsilon}}{T}\;\;,\;\;S_1=\sqrt{\frac{SU}{T}}\;\;,\;\;S_2=\sqrt{\frac{TU}{S}}\;.
\eeq
Then
\beq
T_{a,1}=\frac{i\sqrt{T}\epsilon^{-a/2}}{\sqrt{S U}(1-T)(T-\epsilon)}\(
\frac{T-SU\epsilon^{a+2}}{1-\epsilon}
-
\frac{S T^{a+3}-U\epsilon^{a+2}/T^{a}}{T^2-\epsilon}
\)
\eeq
which for small $\epsilon$ coincides with \eq{Tleading} up to a gauge transformation \eq{gauge} and thus
zero $\epsilon$ limit is the asymptotic limit we consider in the previous section.
We can now easily compute the large $a$ limit of $Y_{\fm_a}$ and $Y_{\fp_a}$:
\beq
\la{bp1}
\lim_{a\to\infty}\frac{\log Y_{\fp_a}(z-i0)}{a}=\log \frac{1}{T}\;\;,\;\;
\lim_{a\to\infty}\frac{\log Y_{\fm_a}(z-i0)}{a}=\log \epsilon\;.
\eeq
From this, using \eq{bp} and \eq{bm}, we fix two of the four unknown function,
\beq
T=\frac{f}{\bar f}\;\;,\;\;\epsilon=\Delta f^2\;.
\eeq
So far we have constructed all $Y_{\fm_a}$ and $Y_{\fp_a}$
in terms of two yet unknown functions $S(z)$ and $U(z)$
in such way that \eq{ie2} and \eq{ie3} are already satisfied.
We have to glue this solution with the solution inside the right and left wings \eq{bb}
parameterized by a third unknown function $A(z)$
and demand the remaining equations (\ref{ne1}-\ref{ne4},\ref{Ynl}) to be satisfied.

\subsubsection{Matching wings}\la{eq:Ysolved}
In the previous section we construct the solution for the upper wing of
$Y$-system (the light gray dots on Fig.2) and \eq{bb} gives the general
solution for the left and right wings $Y_{\fb_s}$.
We parameterized these functions in terms of three unknown functions $S(z),\;U(z),\;A(z)$.
We still have to match these two solutions in the different domains
and find the two remaining fermionic $Y$-functions $Y_{\fF}(z)$ and $Y_{\ff}(z)$.
To fix these five functions we have exactly five remaining equations
(\ref{ne1}-\ref{ne4},\ref{Ynl}).
Excluding $Y_{\fF}$ and $Y_{\ff}$ we get
\beqa
\la{we1}&&\frac{Y_{\fp_2}^2(1+1/Y_{\fp_3})}{1+Y_{\fm_2}}=\frac{F}{(A-1)^2}\;,\\
\la{we2}&&\frac{1+1/Y_{\fp_2}}{1+Y_{\fm_1}}=\frac{1}{A^2}\(\frac{(A-1)^2}{F}-1\)^2\;,\\
\la{we4}&&Y_{\fm_1}^2(1+1/Y_{\fm_2})=\(\frac{(A-1)^2(F-1)}{(A-1)^2-F}\)^2\;,
\eeqa
where $F\equiv Y_{\fF}Y_{\ff}$. $F$ can be expressed in terms of $Y_{\fm_n}$ from \eq{Ynl}:
\beq
F=\la{we3}\frac{1}{f\bar f}\prod_{m=1}^\infty(1+Y_{\fm_m})\;.
\eeq
Notice that the right hand side
of (\ref{we1}-\ref{we3}) depends on $A$ and $F$ only while
the left hand side of these equations does not depend on
the two unknown functions $S$, $U$ and the known functions $f$, $\bar f$ and $\Delta$.
These equations are relatively easy to solve perturbatively in $\Delta$.
For example for $A$ with ${\cal O}(\Delta^4)$
 precision we found
\beqa\nn
A\!&\simeq&
 \frac{1}{\bar{f}-1}+\frac{f}{f-1} +\Delta
    (\bar{f}+f-2 )+\Delta ^2
    (\bar{f}^2+f^2-2 )+\Delta ^3
    (\bar{f}^3+f^3-2 )+\dots\;.
\eeqa
We see that the expansion coefficients are very simple. We can easily sum them up
to get the exact result
\beqa
\nn A&=&
-\frac{1}{\Delta  \bar{f}-1}+\frac{f \bar{f}-1}{(f-1)
    (\bar{f}-1 )}-\frac{1}{f \Delta -1}+\frac{2}{\Delta
   -1}\;,     \\
\nn F&=&
2-\frac{2 (f-1)^2  (\bar{f}-1 )^2}{f (\Delta -1)^3
   \bar{f}}-\frac{(f-1)  (5 f\bar f-3 \bar{f}-3 f+1 )
    (\bar{f}-1 )}{f (\Delta -1)^2 \bar{f}}-\frac{ (\bar{f}+f-2
   f \bar{f} )^2}{f (\Delta -1) \bar{f}}-f \bar{f}\;,
\\
\la{sol} S&=&
\frac{\bar{f}(f-1)^2   (\bar{f}\Delta  -1 )^2}{f
    (\bar{f}-1 )^2 (f \Delta -1)^2}\;,
\\
\nn U&=&\frac{(f-1)^2  (\bar{f}-1 )^2}{f^2 \bar{f}^2 (f \Delta
   -1)^2  (\bar{f}\Delta-1 )^2}
   \;.
\eeqa
The result is quite simple compared to what one may expect to get from
the high degree polynomial equations
and apparently there should exist some more straightforward way to get this result.
It is easy to check that
(\ref{we1},\ref{we2}) and \eq{we4} are indeed satisfied.
To check \eq{we3} one can use \eq{TtoY2}
to get rid of the infinite product
\beq\la{infprod}
F
=\frac{1}{f\bar f}\prod_{m=1}^\infty(1+Y_{\fm_m})
=\frac{1}{f\bar f}\prod_{m=1}^\infty\frac{T_{m,0}^2}{T_{m-1,0}T_{m+1,0}}
=\frac{1}{f\bar f}\frac{T_{1,0}}{T_{0,0}}\lim_{a\to\infty}\frac{T_{a,0}}{T_{a+1,0}}
=\frac{T_{1,0}}{T_{0,0}}\Delta\;,
\eeq
which allows to express the r.h.s. as a rational function of $\epsilon,U,S$ and $T$
so that \eq{sol} can be easily checked. Notice that we found all
$Y$-functions except for the fermionic
ones for which we only presented explicitly the form of their product $F$.
However, using  e.g. (\ref{ne4}) we can easily get $Y_{\fF}$ in terms of
 the other $Y$'s that we just fixed.

We can now plug the functions we just found into
the $Y_{a,s}$ functions to get explicit expressions for {\it{all}}
of these functions in terms of $f$, $\bar f$ and $\Delta$
alone! We recall that these functions are completely fixed
in terms of the Bethe roots (\ref{fs},\ref{DELTA}).
Since the results are not particularly simple, we
present them in Appendix B in {\textsl{Mathematica}} form.
\subsection{Energy and Momentum}
Having all $Y$'s computed we can easily evaluate the energy
of the state, corresponding to a given distribution of roots
from \eq{Energy}.
Using it at strong coupling
\beq
\epsilon_1^{\ph}(z_k)=\frac{x_k^2+1}{x_k^2-1}+{\cal O}\(\frac{1}{g^2}\)\;\;,\;\;
\epsilon_a^{\mir}(z)=-\frac{iaz}{\sqrt{1-z^2}}\;,
\eeq
and applying the trick from the previous section to compute infinite products of
$1+Y_{as}$ used in \eq{infprod} to get
\beq\la{mp0}
e^{{\cal M}_0}\equiv\prod_{a=1}^\infty (1+Y_{\fm_a})^a=\lim_{a\to\infty}\frac{(T_{a,0})^{a+1}}{T_{0,0}(T_{a+1,0})^{a}}
=
\frac{(f \Delta -1)^4  (\bar{f}\Delta  -1 )^4}{(\Delta
   -1)^4    (f \bar{f}\Delta
   -1 )^2 (f^2 \Delta -1 ) (  \bar{f}^2\Delta-1 )}
\;.
\eeq
From where we get the following stunningly simple expression, accurate to all orders in wrapping
\beq\la{energyY0}
E=\sum_{i=1}^M\frac{x_i^2+1}{x_i^2-1}+
\int_{-1}^1
\frac{dz}{2\pi}\frac{z}{\sqrt{1-z^2}}\d_z{\cal M}_0\;.
\eeq
Using the expressions from Appendix B one can see that each separate
term in \eq{mp0} is significantly more complicated then the resulting product! One can easily check \eq{mp0} using explicit expressions for $Y_{\fm_a}$
from Appendix C and expanding both sides of the equality in powers of $\Delta$.

In (\ref{energyY0}), the integration goes over $z\in(-1,1)$ because outside these
region $Y_{\fm_a}$ are strongly suppressed (see \eq{Ymsup}). Notice that the first term is of order $M\sim g\sim \sqrt\lambda$ and contains the classical string energy,
whereas the second is $\sim 1$
and should be a part of the one-loop correction.
The reader may already suspect that the numerator of \eq{mp0} corresponds to $4+4$ fermionic fluctuation modes
whereas the terms in the denominator correspond to $4$ -- modes of $S^5$ and $2+1+1$ --  modes of $AdS_5$.
We make this relation more precise in Sec.\ref{sec:qqs}.
To separate
the classical string energy from the one-loop corrections in the first term of \eq{energyY}
we should find the equation determining positions of $z_i$ with $1/g$ precision.
In the next section we will consider the exact Bethe ansatz equation \eq{BAEex}
in the strong coupling limit.

The total momentum of the state can be computed similarly to
the energy. One should simply replace the
expression for the magnon energy $\epsilon_a$ in \eq{Energy} by the
magnon momentum
\beq\la{mom}
P=\sum_{i=1}^M\frac{x_i}{g(x_i^2-1)}+
\frac{1}{2g}\int_{-1}^1
\frac{dz}{2\pi}\frac{1}{\sqrt{1-z^2}}\d_z{\cal M}_0\;.
\eeq
The natural extension of the cyclicity condition \eq{zeromom0} is
\beq\la{zmom}
P=2\pi m\;.
\eeq
This is an additional constrain and
one should prove its consistency with the other equations.
We will assume \eq{zmom} to be satisfied. 

\subsection{Exact Bethe ansatz equations}\la{ExectBAE}
In the previous section we saw that the exact energy of a given state in the semi-classical limit is given by
\beq\la{energyY}
E=\sum_{i=1}^M\frac{x_i^2+1}{x_i^2-1}+
\int_{-1}^1
\frac{dz}{2\pi}\frac{z}{\sqrt{1-z^2}}\d_z\log \frac{(f \Delta -1)^4  (\bar{f}\Delta  -1 )^4}{(\Delta
   -1)^4    (f \bar{f}\Delta
   -1 )^2 (f^2 \Delta -1 ) (  \bar{f}^2\Delta-1 )}\,\, , \,\,
\eeq
and the functions $f,\bar f,\Delta$ determined uniquely in terms of the Bethe roots
$z_i$ (\ref{fs},\ref{DELTA}).
We still need to find the positions of the roots $z_i$ in order to get the exact energy of the state.
In this section we shall derive that these Bethe equations,
accurate to all orders in wrapping with one-loop precision,
read
\beqa\la{BAEcor}
\nn1&=&-\(\frac{x_k^-}{x_k^+}\)^L\prod_{j=1}^M\frac{x_k^--x_j^+}{x_k^+-x_j^-}\frac{1-1/(x_k^+x_j^-)}{1-1/(x_k^- x_j^+)}
\sigma^2(z_k,z_j)\\
&\times&\exp\[-2\int_{-1}^1 \Big(r(x_k,z){\cal M}_+
-r(1/x_k,z){\cal M}_-
+u(x_k,z){\cal M}_0\Big)dz-iP\]\;,
\eeqa
where
\beqa
e^{{\cal M}_+}=\frac{(f\Delta-1)^2}{(f^2\Delta -1)(f\bar f\Delta-1)}\,\, , \,\,
e^{{\cal M}_-}
=\frac{(\bar f\Delta-1)^2}{(\bar f^2\Delta -1)(f\bar f\Delta-1)}\;,
\eeqa
The factor $\sigma^2(z_k,z_j)$ contains both the leading Arutyunov--Frolov--Staudacher (AFS) \cite{Arutyunov:2004vx}
 and the sub-leading Hernandez--Lopez (HL) phase \cite{Hernandez:2006tk}.
 Initially the AFS phase was designed to give an agreement with classical theory.
Then it was realized that an extra phase is needed in order
 to get an agreement with the semi-classical one-loop string energies \cite{FS1}.
 Basing on the known expressions for the one-loop energies of particular classical solutions
\cite{C2}
 this extra phase
 was found in \cite{Hernandez:2006tk}.
 However, in \cite{SchaferNameki:2006gk} it was shown
 that even with these both dressing factors the Bethe ansatz equations
are
 misses some exponential corrections.
 In \cite{Gromov:2007ky,Gromov:2007cd} the one-loop compatibility of the asymptotical Bethe ansatz
 was proven for a generic
 classical string motion in $AdS_5\times S^5$.
 In \cite{Gromov:2007ky,Gromov:2007cd} it was also noticed that in order to get the agreement
 one should drop some definite exponential in $L/g$ terms.
 In the next section we will work with finite $L/g$
 keeping all the previously dropped terms and show that
 the above equation, obtained from the Y-system, describes accurately the
 one-loop string energies for the $\frak{sl}(2)$ sub-sector.

We now derive the above mentioned equations.
To find the exact positions of the Bethe roots $z_i$
one has to evaluate $Y_{\fm_1}(z_i)$ on the physical real axis (see Fig.3).
The results we obtain in the previous sections are applicable in the domains
where $Y$'s are smooth functions, however close to the Bethe roots one may
expect poles and the approximation  used so far is no longer valid.
\FIGURE[t]{
\epsfig{file=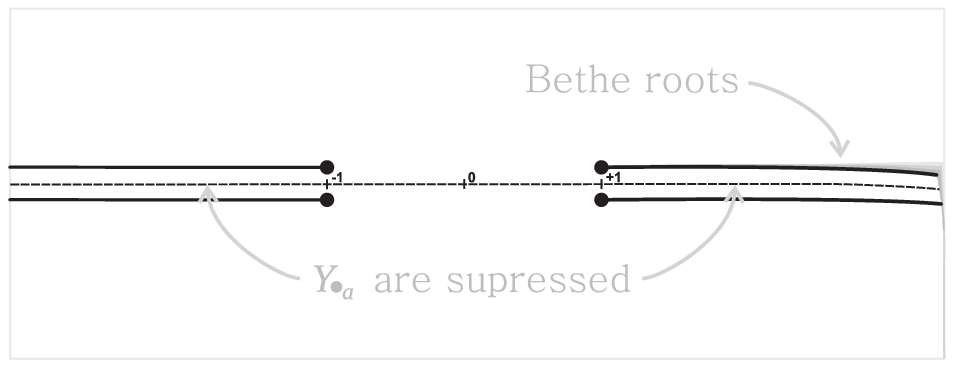,scale=1.4}
\caption{
Branch-cut structure of $Y_{\!{\displaystyle{\bullet}}_{\!1}}$.
Dashed line is a ``mirror" real axis. Between the branch cuts
all $Y_{\!{\displaystyle{\bullet}}_{\!a}}$ are exponentially suppressed at strong coupling
and one can use the asymptotic solution here. To
find the exact positions of the Bethe roots one should analytically
continue $Y_{\!{\displaystyle{\bullet}}_{\!1}}$ under the upper cut to reach a ``physical" real axis.}
}

To get round these difficulties
we use the representation of the $Y$-system
obtained in \cite{Gromov:2009bc}, based on the TBA approach for the ground state
\cite{Bombardelli:2009ns,Gromov:2009bc}.
The equation we need is the integral equation for the middle node
\beq\la{eqmid}
\log Y_{\fm_1}={\cal T}_{1m}*\log (1+Y_{\fm_m})+2{\cal R}^{(10)}\cut\log(1+Y_{\fF})
+
2{\cal R}^{(10)}\cut K_{m-1}*\log(1+Y_{\fp_m})+i\Phi\;,
\eeq
where ${\cal T}_{1m}$ is a kernel containing the dressing phase.
$*$ stands for convolution with integration over real axis,
whereas $\cut$ is a convolution with integration around the cut $(-1,1)$ see
Appendix A for more details.
To use this equation we will need to know
$Y_{\fp_m}$ and $Y_{\fm_m}$ on the whole real axis.
In Sec.\ref{sec:nw} we noticed that for $|z|>1$
the asymptotic solution
from Sec.\ref{sec:nw} can be used since
$Y_{\fm_a}(z)$ are strongly suppressed
for these values of $z$
whereas for $|z|<1$ the solution of the $Y$-system was built above.

We denote by $Y_{as}^0$ the asymptotic solution, constructed
for a given set of {\it exact} Bethe roots $z_i$ in Sec.\ref{sec:bc},
satisfying the exact Bethe equation $Y_{\fm_1}(z_i)=-1$.
The asymptotic solution should satisfy the following integral equation
\beqa
\log Y^0_{\fm_1}&=&2{\cal R}^{(10)}\cut\log(1+Y^0_{\fF})+
2{\cal R}^{(10)}\cut K_{m-1}*\log(1+Y^0_{\fp_m})+i\Phi\;,
\eeqa
we can subtract it from \eq{eqmid} to get
\beqa\nn\la{YmTBA}
\log \frac{Y_{\fm_1}}{Y_{\fm_1}^0}&=&{\cal T}_{1m}*\log (1+Y_{\fm_m})+
2{\cal R}^{(10)}\cut\log\(\frac{1+Y_{\fF}}{1+Y_{\fF}^0}\)+
2{\cal R}^{(10)}\cut K_{m-1}*\log\(\frac{1+Y_{\fp_m}}{1+Y_{\fp_m}^0}\)\;.
\eeqa
As a result of these trick we do not need any more to go outside $-1<z<1$ region in the convolutions,
since the integrands vanish there.
We can now analytically continue $Y_{\fm_1}/Y_{\fm_1}^0$ to the physical real axis where
the Bethe roots are situated (see Fig.3). A similar analytical continuation was
already performed in \cite{Gromov:2009zb}
\beqa\la{YmTBA}
\log \frac{Y_{\fm_1}^\ph}{Y_{\fm_1}^{\ph0}}&=&{\cal T}^{\ph,\mir}_{1m}*\log (1+Y_{\fm_m})+
2{\cal R}^{(10)\ph,\mir}\cut\log\(\frac{1+Y_{\fF}}{1+Y_{\fF}^0}\)\\
\nn&+&
2{\cal R}^{(10)\ph,\mir}\cut K_{m-1}*\log\(\frac{1+Y_{\fp_m}}{1+Y_{\fp_m}^0}\)
+2K_{m-1}(z_k-\tfrac{i}{4g})*\log\(\frac{1+Y_{\fp_m}}{1+Y_{\fp_m}^0}\)\;.
\eeqa
Now we simply have to expand the kernels at large $g$ and
substitute $Y$'s. To expand ${\cal T}^{\ph,\mir}_{1m}$
again can use the AFS dressing phase \eq{AFS}
\beqa
\nn{\cal R}^{(10)\ph,\mir}(z_k,w)&\simeq& r(x_k,w)\;\;,\;\;
{\cal B}^{(10)\ph,\mir}(z_k,w)\simeq r(1/x_k,w)\;,\\
{\cal T}^{\ph,\mir}_{1m}(z_k,w)&\simeq&-m\[2r(x_k,w)+2u(x_k,w)+p(w)\]\;,\\
\nn K_{m}(z_k-w)&\simeq& 
\delta(w-z_k)+m \[r(x_k,w)+ r(1/x_k,w)\]\;,
\eeqa
where we use the following notations
\beqa
\nn r(x,z)&=&\frac{x^2}{x^2-1}\frac{\d_z}{2\pi g}\frac{1}{x-x(z)}\;\;,\;\;
u(x,z)=\frac{x}{x^2-1}\frac{\d_z}{2\pi g}\frac{1}{x^2(z)-1}
\;\;,\;\;
 p(z)= \frac{\d_z}{4\pi i g}\frac{1}{\sqrt{1-z^2}}\;.
\eeqa
After that we can rearrange
the terms in \eq{YmTBA} according to kernels $r$.
Using the following ``magic" products\footnote{To compute
these products we again use $z-i0$ prescription to ensure their convergence.
This prescription is inherited from the TBA equation where the integration
should go slightly below the real axis.}
\beqa
e^{-{\cal M}_+}&\equiv&\frac{1+Y_{\fF}}{1+Y_{\fF}^0}\prod_{m=2}^\infty\(\frac{1+Y_{\fp_m}}{1+Y_{\fp_m}^0}\)^m\prod_{m=1}^\infty \frac{1}{(1+Y_{\fm_m})^m}
=\frac{(f^2\Delta -1)(f\bar f\Delta-1)}{(f\Delta-1)^2}\\
e^{+{\cal M}_-}&\equiv&\frac{1+1/Y_{\ff}^0}{1+1/Y_{\ff}}\prod_{m=2}^\infty\(\frac{1+Y_{\fp_m}}{1+Y_{\fp_m}^0}\)^{m-2}
=\frac{(\bar f\Delta-1)^2}{(\bar f^2\Delta -1)(f\bar f\Delta-1)}\;,
\eeqa
we get the corrected ABA equations (\ref{BAEcor}})
accurate to all orders in wrapping with one-loop precision.
Here we use the expression for the total momentum \eq{mom}.
Notice that the last term in the exponent is irrelevant due to \eq{zmom}.

\subsection{Finite gap solutions}\la{sec:FGS}
In this section we expand the corrected Bethe ansatz equation
\eq{BAEcor}, obtained in the previous section for a particular type
of configurations of roots. Before expanding
 \eq{BAEcor} one should take $\log$ of both sides. Due to the $2\pi i$
 ambiguity of the $\log$ function one should assign an integer mode number
  $n_k$ for each root $x_k$,
  i.e. for each of the $M$ equations we can assign $\log(1)=2\pi i n_k$
  in the left hand side of (\ref{BAEcor})}.
For the finite gap solutions we assume that the set of mod numbers ${n_k}$
contains only a finite number of different integers.
  \FIGURE[ht]{\la{fig:dens}
\epsfig{file=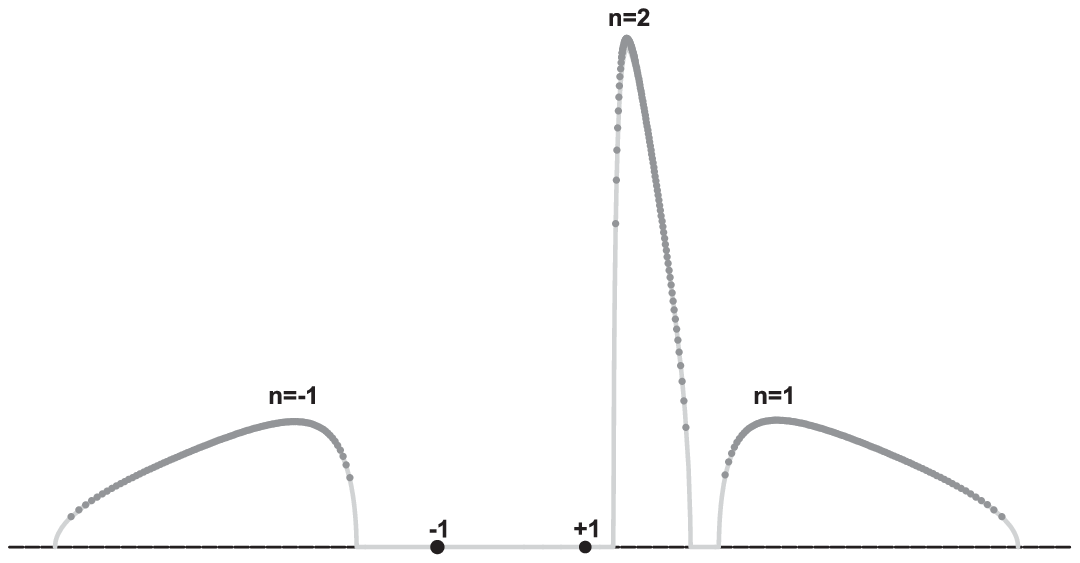,scale=1}
\caption{Density as a function of $z$
for a typical distribution of bethe roots $z_i$ on the real axis.}
} In this limit the Bethe roots $x_k$
are forming dense distributions along some cuts ${\cal C}_n$ on the real axis
with some density $\rho(x)$
so that the number of cuts is equal to the number of distinct mod numbers $n_k$ (see Fig.\ref{fig:dens}).
One can establish a one-to-one correspondence
between such configurations and finite gap classical strings
motions considered in the next section.

Strong coupling expansion of ABA for these configuration was studied intensively \cite{FS4,FS1,BF,Hernandez:2006tk}.
Here we can use the existing results to expand our corrected by wrappings Bethe equation \eq{BAEcor}.
Usually one defines the quasi-momenta through resolvent \eq{ResDef}
\beq
p_{\hat 2}(x)=\frac{\frac{L}{2g}x+G(0)}{x^2-1}+G(x)\;.
\eeq
From the definition of the resolvent \eq{ResDef} we see that $p(x)$ has poles at $x=x_k$.
When the number of roots goes to infinity these poles condense into branch cuts
and we can rewrite the quasi-momenta in terms of the density $\rho$ of the roots
\beq
p_{\hat 2}(x)=\frac{\frac{L}{2g}x+G(0)}{x^2-1}+\sum_n\int_{{\cal C}_n} \frac{\rho(y)}{x-y}\frac{y^2}{y^2-1}dy\;.
\eeq
Equation \eq{BAEcor} gives an integral equation on the density of the roots
\beqa
\nn2\pi n&=&\[p_{\hat 2}(x+i0)+p_{\hat 2}(x-i0)\]+\alpha(x)p'_{\hat 2}(x)\cot p_{\hat 2}(x)+{\cal V}(x)\\
&-&2i\int_{-1}^1 \Big(r(x,z){\cal M}_+
-r(1/x,z){\cal M}_-
+u(x,z){\cal M}_0\Big)dz
\;\;,\;\;x\in {\cal C}_n\;,\la{BAEexp}
\eeqa
where $
\alpha(x)=\frac{x^2}{g(x^2-1)}$.
The second term in \eq{BAEexp} is so-called ``anomaly" term  \cite{FS4,FS1,Beisert:2005di,BF}
and the third therm is the contribution of subleading term in the dressing phase --
the Hernandez-Lopez phase \cite{Hernandez:2006tk}
 \beq
{\cal V}(x)=\alpha(x)\!\!\!\!\!
\sum_{\scriptsize\bea{c}r,s=2\\r\!+\!s\in Odd\eea}^\infty
\!\!\!\!
\frac{1}{\pi}\frac{(r-1)(s-1)}{(s-r)(r+s-2)}\(\frac{Q_r}{x^s}-\frac{Q_s}{x^r}\) \la{HL}\;\;,\;\;
Q_s=\cw\oint\frac{dx}{2\pi i}\frac{G(x)}{x^s}\;.
\eeq
The last line in \eq{BAEexp} incorporates  the finite size effects.
Finally using the standard notations
\beq
p_{\hat 1}(x)=-p_{\hat 2}(1/x)\;\;,\;\;p_{\tilde 2}(x)=\frac{\frac{L}{2g}x+G(0)}{x^2-1}\;
\eeq
we can rewrite ${\cal M}$'s in terms of the quasi-momenta
\beqa
\nn e^{{\cal M}_0}&=&\frac{(1-e^{-ip_{\hat 2}-ip_{\tilde 2}})^4  (1-e^{-ip_{\hat 1}-ip_{\tilde 2}})^4}
{(1-e^{-2i p_{\tilde 2}})^4    (1-e^{-i p_{\hat 1}-i p_{\hat 2}})^2 (1-e^{-2ip_{\hat 2}} ) (1-e^{-2ip_{\hat 1}} )}\;,\\
\la{Masp}e^{{\cal M}_+}&=&\frac{(1-e^{-ip_{\hat 2}-ip_{\tilde 2}})^2}{(1-e^{-2ip_{\hat 2}} )(1-e^{-i p_{\hat 1}-i p_{\hat 2}})}\;,\\
\nn e^{{\cal M}_-}&=&\frac{(1-e^{-ip_{\hat 1}-ip_{\tilde 2}})^2}{(1-e^{-2ip_{\hat 1}} )(1-e^{-i p_{\hat 1}-i p_{\hat 2}})}\;.
\eeqa
In the next section we will see how these structures appear
in the quasi-classical string quantization.

We have found a set of equations which are supposed to correct
the Beisert-Staudacher asymptotic equations with the Beisert-Eden-Staudacher
dressing phase
 in the strong coupling
scaling limit. The latter are known to describe the semi-classical
string spectrum up to exponentially suppressed finite size corrections
as described in the previous section.
This extra corrections which we just derived ought to cure the known
mismatch and correctly incorporate \textit{all} wrapping corrections
to 1-loop precision. In the next section we show that this
turns out to be precisely the case!

\section{Quasi-classical string quantization}\la{sec:qqs}
In this section we review the quasi-classical
quantization method. Then we consider a generic
solution of string equations of motion inside $AdS_3\times S^1$
and compute its one-loop energy.

The one-loop correction to a classical string energy
could be understood as zero point oscillations of
fluctuations around the classical solution. To compute it one
can expand the classical action up to the quadratic order
around the classical solution and
then find the spectrum of oscillation modes.
These modes could be labeled by the mode number $n$,
which tells us how many wavelengthes fit into the string,
and polarization. There are $8+8$ bosonic and fermionic polarizations
which we label by double indices $(ij)$:
\beqa\nn
{\rm Bosonic:\;}&&\;
(\hat 1,\hat 3),\;
(\hat 1,\hat 4),\;
(\hat 2,\hat 3),\;
(\hat 2,\hat 4),\;
(\tilde 1,\tilde 3),\;
(\tilde 1,\tilde 4),\;
(\tilde 2,\tilde 3),\;
(\tilde 2,\tilde 4),\;\\
{\rm Fermionic:\;}&&\;
(\hat 1,\tilde 3),\;
(\hat 1,\tilde 4),\;
(\hat 2,\tilde 3),\;
(\hat 2,\tilde 4),\;
(\tilde 1,\hat 3),\;
(\tilde 1,\hat 4),\;
(\tilde 2,\hat 3),\;
(\tilde 2,\hat 4).\la{eq:in}
\eeqa
We denote the energies of the vibrations  $\Omega_n^{ij}$.
Then the one-loop correction is simply a sum of halves of these fluctuation energies~\cite{Frolov:2002av,Vicedo:2008jy}
\beq\la{Esum}
\delta E_{1-loop}=\frac{1}{2}\sum_{n,(ij)}(-1)^{F_{ij}} \Omega_n^{ij}\;,
\eeq
where $F_{ij}$ is $+1$ for bosonic polarizations and $-1$ for fermionic.

The direct computation of these $\Omega_n^{(ij)}$ is only possible
in the simplest cases \cite{C1,C2}.
For a generic solution it is enormously hard to perform this calculation
starting from the classical action.
The tool which allows to handle the quasi-classical string quantization
efficiently is the algebraic curve technique developed in
\cite{BF,Gromov:2007aq,Gromov:2008ec,Gromov:2008ie}. Below we describe
the construction of the algebraic curve and the method of the quasi-classical calculations.

\subsection{Classical algebraic curve}\la{sec:ac}
The classical equations of motion of the Metsaev-Tseytlin superstring action \cite{Metsaev:1998it}
can be summarized in a compact form as the flatness condition \cite{Bena:2003wd}
\beq
dA-A\wedge A=0\;.
\eeq
 for a connection
$A(\sigma,\tau;x)$ which is a local functional of fields
depending on an arbitrary complex number called the spectral parameter $x$ and taking its values in $\frak{psu}(2,2|4)$. The fact that the classical
equations of motion could be packed into the flatness condition is an indication that the
model is classically integrable. Indeed, we can define the monodromy matrix
\beq
M(x)={\rm Pexp}\oint_\gamma A(x)
\eeq
where $\gamma$ is a loop wrapping the worldsheet cylinder once.
The flatness of the connection ensures path independence of the spectral data of the super $(4+4)\times (4+4)$ matrix $M(x)$.
In particular, the displacement of the whole loop in time direction amounts to a similarity
transformation and we conclude that the eigenvalues of
the monodromy matrix are conserved with time quantities depending on the spectral parameter $x$.
We denote the eigenvalues of $M(x)$ as
\beq
\{e^{i p_{\hat 1}},e^{i p_{\hat 2}},e^{i p_{\hat 3}},e^{i p_{\hat 4}}|e^{i p_{\tilde 1}},e^{ip_{\tilde 2}},e^{ip_{\tilde 3}},e^{ip_{\tilde 4}}\}\;,
\eeq
where $p_{\hat \imath}(x)$ and $p_{\tilde \imath}(x)$ are so-called quasi-momenta
\cite{Kazakov:2004qf,Beisert:2005bm}.
The quasi-momenta contain information about all conserved charges of the theory, in particular the global symmetry charges, including the energy $E$.
The eigenvalues are the roots of the characteristic polynomial and thus they define an $8$-sheet Riemann surface.
In general these sheets are connected by several branch-cuts.
The branch points on this surface are the values of the spectral parameter $x$ where two eigenvalues coincide
and $M(x)$ cannot be diagonalized completely.
Different classical solutions correspond to different algebraic curves. For many calculation
the explicit construction of the classical solution in terms of the initial fields entering into
the Lagrangian is not needed and can be replaced by the corresponding algebraic curve.
For example the energy can be computed as a simple contour integral
\beq\la{Eeq}
E=\sum_{{\cal C}_{ij}}2g\oint_{{\cal C}_{ij}} \frac{dx}{2\pi i}\frac{p_{i}}{x^2}\;.
\eeq

It is always possible to define the quasi-momenta so that they
vanish at  $x\to\infty$. Then, however, the quasi-momenta should jump by a multiple of $2\pi$
when passing through a cut
\beq\la{peq}
\frac{p_{i}(x+i0)+p_{i}(x-i0)}{2}-\frac{p_{j}(x+i0)+p_{j}(x-i0)}{2}=2\pi n\;\;,\;\;x\in{\cal C}\;.
\eeq
The quasi-momenta are restricted by the properties of the
monodromy matrix $M(x)$. Due to super-tracelessness
\beq
p_{\hat 1}+p_{\hat 2}+p_{\hat 3}+p_{\hat 4}=
p_{\tilde 1}+p_{\tilde 2}+p_{\tilde 3}+p_{\tilde 4}\;,
\eeq
and as a consequence of the special properties of $M(x)$ under $x\to 1/x$
transformation one has
\beqa\la{xtoix}
&&p_{\hat 1}(x)=-p_{\hat 2}(1/x)\;\;,\;\;p_{\tilde 1}(x)=-2\pi m-p_{\tilde 2}(1/x)\\
&&p_{\hat 4}(x)=-p_{\hat 3}(1/x)\;\;,\;\;p_{\tilde 4}(x)=+2\pi m-p_{\tilde 3}(1/x)\nn\;,
\eeqa
where $m$ is an integer winding number.

There are also infinitely many points where two eigenvalues coincide but, nevertheless,  the matrix
$M(x)$ can be diagonalized. The two quasi-momenta $p_i$ and $p_j$ corresponding to the coincident eigenvalues
have no singularity and differs by $2\pi n$.
 One can perturb the curve by opening a small cut connecting
the intersecting sheets of the surface at these points.
We label these points by an integer $n$ and a couple of indices $(ij)$
\beq\la{xdef}
p_i(x_n^{ij})-p_j(x_n^{ij})=2\pi n\;.
\eeq

One of the nice features of the algebraic curve is the simplicity
of visualization of the action variables
of this classical integrable theory.
They are the contour integrals around the branch cuts
\beq\la{Sij}
\frac{g}{2\pi i}\oint_{{\cal C}}\(1-\frac{1}{x^2}\)p_i(x)dx\;.
\eeq
In the standard quasi-classical quantization procedure one should assume them to be integers.

\subsubsection{Algebraic curve for $\frak{sl}(2)$ subsector}
The algebraic curve for the string in $AdS_3\times S^1$
was constructed in \cite{Kazakov:2004nh}. In the general framework reviewed in the previous section
this sector corresponds to the cuts connecting $p_{\hat 2}$
with $p_{\hat 3}$ outside the unit circle centered at the origin.
Automatically, due to the  $x\to 1/x$ symmetry \eq{xtoix}, we will have
reflected cuts
connecting $p_{\hat 1}$ with $p_{\hat 4}$ inside the unit circle.
One can easily build the spectral representation
for the quasi-momenta~\cite{Kazakov:2004nh}
\beqa\la{eq:p}
&&p_{\hat 2}(x)=-p_{\hat 3}(x)=
-p_{\hat 1}(1/x)=p_{\hat 4}(1/x)=\frac{\frac{L}{2g}x+G(0)}{x^2-1}+G(x)\;,\\
&&p_{\tilde 2}(x)=-p_{\tilde 3}(x)=
p_{\tilde 1}(x)=-p_{\tilde 4}(x)=\frac{\frac{L}{2g}x+G(0)}{x^2-1}\;,
\eeqa
where $G(x)=\int_{{\cal C}} \frac{\rho(y)}{x-y}\frac{y^2}{y^2-1}dy$.
We see that these quasi-momenta are
exactly those of Sec.\ref{sec:FGS}. The action variables
\eq{Sij} count the number of Bethe roots constituting the cut.
In this way one established the map between classical solutions and
the finite gap configurations of Bethe roots~\cite{Kazakov:2004nh}.

The equation \eq{peq} for the $\frak{sl}(2)$ subsector becomes
\beq\la{eq:sl2}
p_{\hat 2}(x+i0)+p_{\hat 2}(x-i0)=2\pi n\;\;,\;\;x\in{\cal C}\;,
\eeq
which is now an integral equation for the discontinuity $\rho(x)$.

\subsection{Quasi-classical corrections from algebraic curve}\la{sec:QC}
Using the algebraic curve it is also possible to find the spectrum of the fluctuations $\Omega_n^{(ij)}$
around an arbitrary classical solution
using the techniques developed in \cite{BF,Gromov:2007ky,Gromov:2007cd,Gromov:2007aq,Gromov:2008ec,Gromov:2008ie}.
The perturbations of the given classical solution
are reflected in the algebraic curve by extra cuts.
The small cuts could only appear in the special points of the curve
given by \eq{xdef}. The perturbed quasi-momenta differ from the non-perturbed ones
by a small amount $\delta p_i(x)$. The minimal size of the cut
is restricted in the quasi-classically quantized theory
by the condition that the contour integral around this new cut
\eq{Sij} is integer.

From far away the branch points almost merge and the cut looks like
a pole with a tiny residue
\beq\la{dpi}
\delta p_i(x)\sim \frac{\alpha(x)}{x-x_n^{ij}}\;\;,\;\;\alpha(x)=\frac{x^2}{g(x^2-1)}\;,
\eeq
such that \eq{Sij} counts a single quantum. We see that for given $n$ and $(ij)$
the perturbation of the quasi-momenta is pretty much restricted
and one can compute the energy shift due to this fluctuation.
This gives precisely $\Omega_n^{ij}$. In all details this technique
is described in \cite{Gromov:2007aq,Gromov:2008ec,Gromov:2008ie,Gromov:2009zz} (see also \cite{Vicedo:2008jy}).

$\Omega_n^{ij}$
has two contributions different by their nature. Firstly,
the extra small cut carries its own energy as we can see from \eq{Eeq}
\beq\la{om0}
{\Omega^{0}_n}^{ij}={\omega}(x_n^{ij})\;\;,\;\;\omega(x)\equiv\frac{2}{x^2-1}\;,
\eeq
\FIGURE[t]{\la{fig:flac}
\epsfig{file=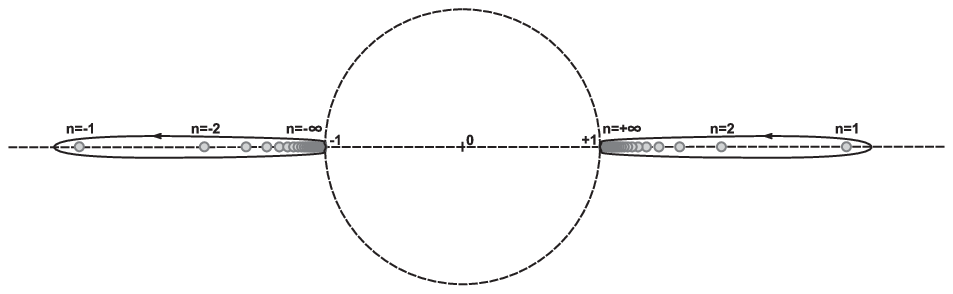,scale=1.4}
\caption{
Complex plane of the spectral parameter $x$. Positions of the fluctuations $x_n^{ij}$
are shown by the gray dots.}
}
secondly,
it deforms others cuts, changing thus their contribution.
This second contribution we study in the next section.
Let us now see the effect of \eq{om0} on the one-loop shift \eq{Esum}.
We have to compute the following sum
\beq\la{sumE}
\delta E_{1-loop}^{0}=\sum_n\sum_{(ij)} (-1)^{F_{ij}}\omega\(x_n^{ij}\)\;,
\eeq
where $x_n^{ij}$ should be found from \eq{xdef}. We rewrite this sum over $n$ as an integral
\beq
\delta E_{1-loop}^{0}=\sum_{(ij)}(-1)^{F_{ij}}\ccw\oint \frac{dn}{4i}\cot(\pi n)\omega\(x_n^{ij}\)\;,
\eeq
where the contour encircles the real axis. Next, for each polarization $(ij)$ we change the integration variable
from $n$ to $x$ via \eq{xdef}.
The integration over $n$ maps to contours which encircle the fluctuation positions
$x_n^{ij}$ located outside the unit circle ${\mathbb U}$.
Then we can deform this contour in the $x$ plane to get an integral over the
unit circle, centered at the origin (see Fig.\ref{fig:flac})\footnote{For each $(ij)$ the sum in $n$ in \eq{sumE} is divergent and one should
carefully treat the large $n$'s.
The one-loop shift written as an integral \eq{eq:dE} corresponds to a particular
prescription of the large $n$ regularization which is analyzed in details in \cite{Gromov:2007cd}.
Whereas from the algebraic curve point of view this prescription
is absolutely reasonable from the worldsheet action point of view
the complete prove is still missing. We should notice however that
this question is not a particularity of the finite size effects, it is rather
related to the question of the validity of the ABA at the one-loop level.}
\beq\la{eq:dE}
\delta E_{1-loop}^{0}=\sum_{(ij)}(-1)^{F_{ij}}\cw\oint_{{\mathbb U}} \frac{dx}{4i}\frac{p_i'-p_j'}{2\pi}\cot\(\frac{p_i-p_j}{2}\)\omega\(x\)
=\cw\oint_{{\mathbb U}^+} \frac{dx}{2\pi i}
\omega\(x\)\d_x {\cal N}_0
\;,
\eeq
where ${\mathbb U}^+$ is the upper half of the unit circle and
\beq\la{eq:N0}
e^{{\cal N}_0}=\prod_{{(ij)}}\(1-e^{-ip_i+ip_j}\)^{F_{ij}}\;.
\eeq
We use that
\beq
\sum_{(ij)}(-1)^{F_{ij}}(p_i-p_j)=0\;.
\eeq
The product in \eq{eq:N0} goes over
all $8+8$ polarizations listed in \eq{eq:in}.
Notice that from \eq{eq:p} $p_i-p_j\sim L/g$
and the integral \eq{eq:N0} is exponentially suppressed for $L/g$
large. This kind of terms are not captured by the ABA and as a result the ABA
can be only used when $L/g$ is sufficiently large.

To our deep satisfaction we notice that
for the $\frak{sl}(2)$ subsector ${\cal N}_0=-{\cal M}_0$ from \eq{Masp}!
Moreover by changing the integration variable to $z=\tfrac{1}{2}\(x+1/x\)$
we map the integration contour to $[-1,1]$ segment of the real axis
and \eq{eq:dE} matches precisely with the second term of the expression
for the energy obtained from $Y$-system \eq{energyY}!
In the next section we show how the corrected Bethe equation
\eq{BAEexp} arises from the quasi-classical quantization.

\subsection{Back-reaction}
So far only the direct contribution
of the virtual sea of the fluctuations was  computed.
We have to take into account the back-reaction --
the deformation of the quasi-momenta close to the
cuts of the initial non-perturbed classical curve.
In \cite{Gromov:2007cd} such deformations were
considered dropping exponentially suppressed finite
size corrections. This allowed for a precise derivation
of the HL correction to the AFS asymptotic Bethe equations.
Here we will keep all exponentially suppressed terms
since we want to derive a set of exact integral equations.
We
split $p_{\hat 2}$ into the part containing
all the small virtual cuts $V_{\hat 2}$
and the smooth part $p_{\hat 2}^{\rm br}$.
To write down $V_{\hat 2}$ one should take into
account $x\to 1/x$ symmetry and some further analyticity constraints
such as poles at $x=\pm 1$.
The basic rule is that each fluctuation $\Omega_{n}^{ij}$
contributes as a pole at $x=x_{n}^{ij}$ with the residue
$\alpha(x)$ on the corresponding
sheets $i$ and $j$ and also by a pole at $1/x_{n}^{ij}$ due to the
constraint \eq{xtoix} (see \cite{Gromov:2007cd} for more details)\footnote{compare with equation 20 in \cite{Gromov:2007cd}.}
\beq\la{eq:pot}
V_{\hat 2}=
\sum_n\(\frac{\alpha(x)}{x}\sum_{(ij)} \frac{(-1)^{F_{ij}}}{(x_n^{ij})^2-1}
+\alpha(x)\sum_{(\hat 2 j)} \frac{(-1)^{F_{ij}}}{x-x_n^{ij}}
-\alpha(1/x)\sum_{(\hat 1 j)} \frac{(-1)^{F_{ij}}}{1/x-x_n^{ij}}
\)\;.
\eeq
The second sum goes over all fluctuations starting at $p_{\hat 2}$
\beq\nn
{\rm Bosonic:\;}\;\;\;
(\hat 2,\hat 3),\;
(\hat 2,\hat 4),\;\;\;\;
{\rm Fermionic:\;}\;\;\;
(\hat 2,\tilde 3),\;
(\hat 2,\tilde 4),\la{eq:in2}
\eeq
and in the last term, corresponding to the reflected poles,
the sum goes over all fluctuations starting at $p_{\hat 1}$.

Now we should use \eq{eq:sl2} to find the discontinuity of $p^{\rm br}_{\hat 2}$
\beq
p^{\rm br}_{\hat 2}(x+i0)+p^{\rm br}_{\hat 2}(x-i0)+2V_{\hat 2}(x)=2\pi n\;\;,\;\;y\in{\cal C}\;.
\eeq
We can convert the sum over $n$ in \eq{eq:pot}
into the integral over $x$ -- precisely like we did with the energy
and then deform the contour to the unit circle
\beq
V_{\hat 2}=\!\!
\cw\oint_{{\mathbb U}^+}\! \frac{dy}{2\pi i}
\(\frac{\alpha(x)}{x}\frac{
\d_y {\cal N}_0}{y^2-1}
+\alpha(x)\frac{\d_y {\cal N}_+}{x-y}
-\alpha(1/x) \frac{\d_y {\cal N}_-}{1/x-y}
\)
+\frac{\alpha(x)}{2}p_{\hat 2}'\cot p_{\hat 2}
+\frac{1}{2}{\cal V}(x).
\eeq
There is one important difference that now there is an extra pole at $x=y$
caught when deforming the contour to the unit circle given rise to the second term (see \cite{Gromov:2007ky}).
We denote
\beq
e^{{\cal N}_+}=\prod_{{(\hat 2j)}}\(1-e^{-ip_{\hat 2}+ip_j}\)^{F_{\hat 2j}}\;\;,\;\;
e^{{\cal N}_-}=\prod_{{(\hat 1j)}}\(1-e^{-ip_{\hat 1}+ip_j}\)^{F_{\hat 1j}}
\eeq
and
\beq
{\cal V}(x)=\cw\oint_{{\mathbb U}^+} \frac{dy}{2\pi}
\(\frac{\alpha(x)}{x-y}
- \frac{\alpha(1/x)}{1/x-y}
\)\d_y(p_{\hat 1}+p_{\hat 2}-p_{\tilde 1}-p_{\tilde 2})\;.
\eeq
taking into account that $\d_y(p_{\hat 1}+p_{\hat 2}-p_{\tilde 1}-p_{\tilde 2})=\d_y(G(y)-G(1/y))$
one can see that this coincides precisely with the contribution
of the Hernandez-Lopez phase~\cite{Gromov:2007cd} in the $Y$-system analysis \eq{HL}.
For $\frak{sl}(2)$ we again have ${\cal N}_{\pm}=-{\cal M}_{\pm}$
and after change of the integration variable to $z=\tfrac{1}{2}(x+1/x)$
we get precisely the equation obtained in the $Y$-system framework \eq{BAEexp}.
Thus we established the match of these two completely different approaches
at the level of equations.

\section{Summary and future directions}

In this paper we studied the finite size effects at strong coupling for strings in $AdS^3\times S_1$.
We attacked the problem from two directions -- from the quasi-classical string
quantization using the algebraic curve techniques
\cite{Gromov:2007aq,Gromov:2008ec,Gromov:2008ie,Gromov:2009zz} and from the
recently conjectured $Y$-system \cite{Gromov:2009tv}. We found the same result
in both cases thus providing
a very nontrivial test of the latter.
We also
derived the corrected expression
for the energy of \eq{energyY}
\beq\la{energyY6}
E=\sum_{i=1}^M\frac{x_i^2+1}{x_i^2-1}+
\int_{-1}^1
\frac{dz}{2\pi}\frac{z}{\sqrt{1-z^2}}\d_z{\cal M}_0\;\;,\;\;e^{{\cal M}_0}= \frac{(f \Delta -1)^4  (\bar{f}\Delta  -1 )^4}{(\Delta
   -1)^4    (f \bar{f}\Delta
   -1 )^2 (f^2 \Delta -1 ) (  \bar{f}^2\Delta-1 )}\,,
\eeq
where $f$ and $\bar f$ \eq{fs} are some simple functions of the Bethe roots
$x_i$ and $\Delta$ \eq{DELTA}
is the exponential wrapping parameter.
The last integral term is responsible for the finite size effects
and vanishes in the large volume limit.
We also found that the Bethe roots should satisfy the corrected Bethe equation
\beqa\la{BAEcor6}
\nn-1&=&\(\frac{x_k^-}{x_k^+}\)^L\prod_{j=1}^M\frac{x_k^--x_j^+}{x_k^+-x_j^-}\frac{1-1/(x_k^+x_j^-)}{1-1/(x_k^- x_j^+)}
\sigma^2(z_k,z_j)\\
&\times&\exp\[-2\int_{-1}^1 \Big(r(x_k,z){\cal M}_+
-r(1/x_k,z){\cal M}_-
+u(x_k,z){\cal M}_0\Big)dz\]\;,
\eeqa
where ${\cal M}_+$ and ${\cal M}_-$ contain all exponential wrapping corrections
\beq
e^{{\cal M}_+}
=\frac{(f\Delta-1)^2}{(f^2\Delta -1)(f\bar f\Delta-1)}\;\;,\;\;
e^{{\cal M}_-}
=\frac{(\bar f\Delta-1)^2}{(\bar f^2\Delta -1)(f\bar f\Delta-1)}\;.
\eeq

There are many interesting directions which would be worth exploring:
\begin{itemize}
\item It would be interesting to make a more direct analysis
by solving our corrected equations \eq{BAEcor6} for some simple configuration
of roots  and comparing the solution with the sum of fluctuation energies
for the corresponding classical solution. For example, it would be very
nice to repeat the analysis of \cite{SchaferNameki:2006gk} for circular
strings in $AdS_3$ using the corrected Bethe equations.
\item
It would be also interesting to compare the corrected Bethe
equations (\ref{BAEcor6}) with the conjectured generalized L\"{u}scher
formula \cite{Bajnok:2008qj} in the strong coupling scaling limit.
\item
In this paper we focused on strings moving in $AdS^3 \times S_1$.
From the Y-system point of view this is an important simplification
because the excited states integral equations
are only available for this sector\cite{Gromov:2009bc}. On the other hand, from the string
semi-classics point of view, following \cite{Gromov:2007ky,Gromov:2007cd}, the
derivation of the corrected Bethe equations would be a straightforward
task. It would be very interesting to perform this generalization and
to use it as a guiding principle to construct the Y-system integral
equations for \textit{any} excited state.
\item
It would also be very important to consider analytically some
states which cannot be treated bythe scaling limit -- like Konishi state.
\item The $Y$-system conjectured in \cite{Gromov:2009tv} for
 Aharony-Bergman-Jafferis-Maldacena~\cite{ABJM} theory
recently was supported from the $4$-loop perturbation theory
\cite{Minahan:2009aq} at weak coupling.
It would be also interesting to make
some strong coupling test of this conjecture.
\item Related to finite size corrections but at weak coupling
one should reproduce the results of \cite{Bajnok:2009vm} from the $Y$-system set of equations.
\item Finally one can try to generalize the approach used here to solve
the $Y$-system at finite coupling by bringing it to a couple
of integral equations like in \cite{Gromov:2008gj}
(see \cite{Tsuboi:2009ud} for some first steps).

\end{itemize}

In short, there are many interesting open problems to address
related to the exact computation of the AdS/CFT planar spectrum
and many simplifications are to be expected. We are getting closer and closer to
finding the exact solution to a four dimensional superconformal
gauge theory for the very first time.
The methods developed here could be also useful
for a wide range of integrable theories.
The quasi-classical quantization probes the theories
at finite volume and provides an important information
about hidden structures, such as $Y$-systems.

\section*{Acknowledgments}

The work was partly supported by the German Science Foundation (DFG) under
the Collaborative Research Center (SFB) 676 and RFFI project grant 06-02-16786.
We thank
D.~Serban and A.~Tseytlin for important remarks  on the manuscript,
V.~Kazakov and P.~Vieira for the collaboration on the initial stage of this project
and many useful discussions and F.~Levkovich-Maslyuk for careful reading of the manuscript
 and useful comments.
\appendix
\section{Notations}
There are two distinct possibilities to define $x(z)$ which is solution to $x+1/x=2z$
\beq\label{two branches}
x^{\ph}(z)\equiv z+\sqrt{z-1}\sqrt{z+1}\;\;,\;\;
x^{\mir}(z)\equiv z+i\sqrt{1-z^2}\;.
\eeq
By default we always choose $x=x^\mir$.
These two functions coincide above the real axis and have the following properties under complex
conjugation
\beq
\overline \px=\px\;\;,\;\;\overline  \mx=1/\mx\;.
\eeq
We also use the notation for the Bethe roots
\beq
x_j=x^{\ph}(z_j)\;\;,\;\;
x^\pm_j=x^{\ph}(z_j\pm\tfrac{i}{4g})\;.
\eeq
The single magnon energy and momentum are
\beq
\epsilon_a(z)=a+\frac{2ig}{x(z+\tfrac{ia}{4g})}-\frac{2ig}{x(z-\tfrac{ia}{4g})}\;\;,\;\;
\pi_a(z)=\frac{1}{i}\log\frac{x(z+\tfrac{ia}{4g})}{x(z-\tfrac{ia}{4g})}\;,
\eeq
depending on which $x(z)$ we are using it could be denoted
$\epsilon_a^\ph(z)$ or $\epsilon_a^\mir(z)$.

The kernels we are using in the integral equations are defined as follows\footnote{We use rescaled kernels compare to \cite{Gromov:2009bc}: ${\cal K}^{\rm new}(z,w)=2g {\cal K}^{\rm old}(2gz,2gw)$.}
\beqa
K_{n}(z)&\equiv& \frac{4g n}{\pi(n^2+16g^2 z^2)}\;,\nn\\
K_{nm}(z)&\equiv&
\sum_{k_1=\frac{1-n}2}^{\frac{n-1}2}
\sum_{k_2=\frac{1-m}2}^{\frac{m-1}2}K_{2}\(z+i\tfrac{k_1+k_2}{2g}\)\;,\nn\\
K_{nm}^{\neq}(z)&\equiv&
\sum_{k_1=\frac{1-n}2}^{\frac{n-1}2}
\sum_{k_2=\frac{1-m}2}^{\frac{m-1}2}K_{1}\(z+i\tfrac{k_1+ k_2}{2g}\)\;,\\
 {\cal R}^{(nm)}(z,w)&\equiv&
\frac{\d_w}{2\pi i}\log \frac{\mx(z+\tfrac{in}{4g})-\mx(w-\tfrac{im}{4g})}{\mx(z-\tfrac{in}{4g})-\mx(w+\tfrac{im}{4g})}
+\frac{\d_w}{4\pi i}\log \frac{\mx(w+\tfrac{im}{4g})}{\mx(w-\tfrac{im}{4g})}
\;\;,\;\;\nn\\ \nn
 {\cal B}^{(nm)}(z,w)&\equiv&
\frac{\d_w}{2\pi i}\log \frac{1/\mx(z+\tfrac{in}{4g})-\mx(w-\tfrac{im}{4g})}{1/\mx(z-\tfrac{in}{4g})-\mx(w+\tfrac{im}{4g})}
+\frac{\d_w}{4\pi i}\log \frac{\mx(w+\tfrac{im}{4g})}{\mx(w-\tfrac{im}{4g})}\;, \\
{\cal M}_{nm}&\equiv& K_{n-1}\cut {\cal R}^{(0m)}+K^{\neq}_{n-1,m-1}\;,\nn\\
{\cal N}_{nm}&\equiv&{\cal R}^{(n0)}\cut K_{m-1}+K^{\neq}_{n-1,m-1}\;, \nn\\
\eeqa
There are two types of convolutions $*$ and $\cut$. The first corresponds
to the usual integration along whole real axis whereas the second one
is a convolution a long a path going from $-1$ to $1$ and then
back on another sheet e.g.
\beq
{\cal R}^{(n0)}\cut\log(1+Y_{\fF})=
\int_{-1}^{1} dz \left[ {\cal R}^{(n0)}\log(1+Y_{\fF})-{\cal B}^{(n0)}\log(1+1/Y_{\ff})
\right] \nn
\eeq
where $1/{Y_{\ff}}$
is the analytical continuation of $Y_{\fF}$   across the cut
$u\in (-\infty,-1)\cup(1,+\infty)$.
The kernel ${\cal T}_{1m}(z,w)$ is defined in the following way.
For $\IM z>\frac{in}{4g}$ and $\IM w>\frac{im}{4g}$
we define it using the usual Beisert-Eden-Staudacher dressing factor $\sigma$~\cite{Beisert:2006ez}:
\beqa
\nn{\cal T}_{nm}(z,w)&=&\frac{2}{2\pi i}\frac{d}{dw}\log\sigma\(x(z+\tfrac{in}{4g}),x(z-\tfrac{in}{4g}),x(w+\tfrac{im}{4g}),x(w-\tfrac{im}{4g})\)
\\&+&\sum_{k=-\frac{n-1}{2}}^{\frac{n-1}{2}}\({\cal B}^{(1m)}(z+\tfrac{ik}{2g},w)-{\cal R}^{(1m)}(z+\tfrac{ik}{2g},w)\)\;.
\eeqa
The function ${\cal T}_{1m}(z,w)$ has the branch-points at $z=\pm 1 +\frac{in}{4g}$
and $w=\pm 1+\frac{im}{4g}$. One should analytically continue between them in $z$ and $w$.
Defined in this way function has four branch-cuts going to infinity in $z$ variable
starting at $z=\pm 1 \pm\frac{in}{4g}$ and four branch-cuts going to infinity in $w$ variable
starting at $w=\pm 1 \pm\frac{im}{4g}$ \cite{Arutyunov:2009kf,Gromov:2009bc}.

In \eq{YmTBA} we also use the notation ${\cal R}^{(10)\ph,\mir}$ and ${\cal T}^{\ph,\mir}_{1m}$,
which means that one should take ${\cal R}^{(10)}(z,w)$ (or ${\cal T}_{1m}(z,w)$) and then analytically
continue it in the first argument along a path going around the branch point $z=1+\frac{i}{4g}$.
For ${\cal R}^{(10)}(z,w)$ it simply results in the replacement $x(z\pm \frac{i}{4g})\to x^\ph(z\pm \frac{i}{4g})$.

In the main text we also use the following generalized Baxter polynomials
\beq
R^{(\pm)}(z)=\prod_{j=1}^M\frac{\mx(z)-x^\ph(z_j\mp \tfrac{i}{4g})}{\sqrt{x^\ph(z_j\mp \tfrac{i}{4g})}}\;\;,\;\;
B^{(\pm)}(z)=\prod_{j=1}^M\frac{1/\mx(z)-x^\ph(z_j\mp \tfrac{i}{4g})}{\sqrt{x^\ph(z_j\mp \tfrac{i}{4g})}}\;.
\eeq
They are complex
conjugates one of another $ B^\pm(z)=\overline{R^\mp(z)}$.

\section{Explicit expressions for $Y$'s}
One can copy the expressions below directly to \textsl{Mathematica}
from .pdf.
Here we denote $\verb"d"=\Delta,\;\verb"fb"=\bar f,\;\verb"Ym[a_]"=Y_{\fm_a},\;\verb"Yp[a_]"=Y_{\fp_a},\;
\verb"Yb[s_]"=Y_{\fb_s}$\\
\\
\verb"sbf={e->f^2 d, T->f/fb,"\\
\verb"S->((f-1)^2 fb(d fb-1)^2)/(f(fb-1)^2(d f-1)^2),"\\
\verb"U->((f-1)^2(fb-1)^2)/(f^2 fb^2(d f-1)^2(d fb-1)^2),"\\
\verb"A->1/(1-d f)+1/(1-d fb)+2/(d-1)+f/(f-1)+1/(fb-1),"\\
\verb"F->-((2(f-1)^2(fb-1)^2)/((d-1)^3 f fb))-(f+fb-2 f fb)^2/((d-1)f fb)"\\
\verb"  -((f-1)(fb-1)((5 f-3)fb-3 f+1))/((d-1)^2 f fb)-f fb+2};"\\
\\
\verb"Ym[a_]=(U S(T-1)^2 T e^a(e-T)^2(T^a(e-T^2)(U S e^(a+2)-T)-(e-1)U e^(a+2)+"\\
\verb"(e-1)S T^(2a) T^3)^2)/((-(2 U T e^(a+1)(T-S T^a)(S T^a T-1))"\\
\verb"+T^2(S T^(2a)(T-1)^2-U T e^a(S T^a-1)^2)+U e^(a+2)(U S(T-1)^2 e^a"\\
\verb"-T(S T^a-1)^2))(-(U T e^(a+4)(S T^a T^2-1)^2)-U T^3 e^(a+2)(S T^a T^2-1)^2"\\
\verb"+2 U T^2 e^(a+3)(S T^a T-1)(S T^a T^3-1)+U^2 S(T-1)^2 e^(2 a+6)"\\
\verb"+S T^(2a)(T-1)^2 T^6))/.sbf;"\\
\\
\verb"Yp[a_]=-(((e-1)(e-T^2) T^a(-(U T e^(a+3)(S T^(a+1)-1)^2)-U T^3 e^(a+1)"\\
\verb"(S T^(a+1)-1)^2+2 U T^2 e^(a+2)(S T^a-1)(S T^(a+2)-1)+U^2 S(T-1)^2"\\
\verb"e^(2 a+4)+S(T-1)^2 T^(2 a+4)))/(((e-T^2) T^(a+1)(U S e^(a+3)-T)-(e-1) U"\\
\verb"e^(a+3) +(e-1) S T^(2 a+5))(U e^(a+2)(T-S T^a)+e T(S T^(a+1)-1)"\\
\verb"(U e^a-T^a)+T^(a+2)(S T^a-T))))/.sbf;"\\
\\
\verb"Yb[s_]=(s-A)^2-1/.sbf;"\\
\\
\verb"Y22=(d^2 f^4(fb-1)^2(d fb-1)^2- d f^3(d(d+1)fb^2-4 d fb+d+1)(fb(d(2 fb-3)"\\
\verb"-1)+2)+f^2(d(d(d(d+4)+1)fb^4-8 d(2 d+1)fb^3+2(d(11-(d-7)d)+1)fb^2"\\
\verb"-8(2 d+1)fb+d+4)+1)-f(d(d+1)fb^2-4 d fb+d+1)(fb(d(2 fb-3)-1)+2)+(fb-1)^2"\\
\verb"(d fb-1)^2)/((d-1)f fb(d fb^2((d-3)d f^2-2(d-2)d f+d+2 f-3)-fb(d(2((d-2)"\\
\verb"d-1)f^2-3(d-1)d f+2 d+7 f-4)+f)+(d-3)d f^2- 2(d-2)d f+d+2 f+2 fb-3));"\\
\\
\verb"Y11=F/Y22/.sbf;"



\begin{thebibliography}{99}

\bibitem{Maldacena:1997re}
  J.~M.~Maldacena,
  \textit{``The large N limit of superconformal field theories and supergravity,''}
  Adv.\ Theor.\ Math.\ Phys.\  {\bf 2} (1998) 231
  [Int.\ J.\ Theor.\ Phys.\  {\bf 38} (1999) 1113]
  $\diamondsuit$
  E.~Witten,
  \textit{``Anti-de Sitter space and holography,''}
  Adv.\ Theor.\ Math.\ Phys.\  {\bf 2} (1998) 253


\bibitem{GKP}
  S.~S.~Gubser, I.~R.~Klebanov and A.~M.~Polyakov,
  \textit{``Gauge theory correlators from non-critical string theory,''}
   Phys.\ Lett.\  B {\bf 428} (1998) 105


\bibitem{Minahan:2002ve}
  J.~A.~Minahan and K.~Zarembo,
  \textit{``The Bethe-ansatz for N = 4 super Yang-Mills,''}
  JHEP {\bf 0303} (2003) 013
$\diamondsuit$
  N.~Beisert, C.~Kristjansen and M.~Staudacher,
  \textit{``The dilatation operator of N = 4 super Yang-Mills theory,''}
  Nucl.\ Phys.\  B {\bf 664} (2003) 131
$\diamondsuit$
  R.~A.~Janik,
  \textit{``The $AdS_5 \times S^5$ superstring worldsheet S-matrix and crossing  symmetry,''}
  Phys.\ Rev.\  D {\bf 73}, 086006 (2006)
$\diamondsuit$
  M.~Staudacher,
  \textit{``The factorized S-matrix of CFT/AdS,''}
  JHEP {\bf 0505}, 054 (2005)
$\diamondsuit$
  N.~Beisert,
  \textit{``The $su(2|2)$ dynamic S-matrix,''}
  Adv.\ Theor.\ Math.\ Phys.\  {\bf 12}, 945 (2008)
  [arXiv:hep-th/0511082].
$\diamondsuit$
  N.~Beisert, R.~Hernandez and E.~Lopez,
  \textit{``A crossing-symmetric phase for $AdS_5 \times S^5$ strings,''}
  JHEP {\bf 0611} (2006) 070

\bibitem{Bena:2003wd}
  I.~Bena, J.~Polchinski and R.~Roiban,
  \textit{``Hidden symmetries of the $AdS_5 \times S^5$ superstring,''}
  Phys.\ Rev.\  D {\bf 69} (2004) 046002


\bibitem{Kazakov:2004qf}
  V.~A.~Kazakov, A.~Marshakov, J.~A.~Minahan and K.~Zarembo,
  \textit{``Classical / quantum integrability in AdS/CFT,''}
  JHEP {\bf 0405}, 024 (2004)

\bibitem{Arutyunov:2004vx}
G.~Arutyunov, S.~Frolov and M.~Staudacher, \textit{``Bethe ansatz for quantum
  strings"},
  {\em JHEP} {\bf 10} (2004) 016.


\bibitem{Beisert:2005bm}
  N.~Beisert, V.~A.~Kazakov, K.~Sakai and K.~Zarembo,
  \textit{``The algebraic curve of classical superstrings on $AdS_5 \times S^5$,''}
  Commun.\ Math.\ Phys.\  {\bf 263}, 659 (2006)

\bibitem{Gromov:2009zb}
  N.~Gromov, V.~Kazakov and P.~Vieira,
  \textit{``Exact AdS/CFT spectrum: Konishi dimension at any coupling,''}
  arXiv:0906.4240 .


\bibitem{Gromov:2009tv}
  N.~Gromov, V.~Kazakov and P.~Vieira,
\textit{``Integrability for the Full Spectrum of Planar AdS/CFT,''}
  arXiv:0901.3753 .

\bibitem{Beisert:2006ez}
  N.~Beisert, B.~Eden and M.~Staudacher,
\textit{``Transcendentality and crossing,''}
  J.\ Stat.\ Mech.\  {\bf 0701} (2007) P021
  [arXiv:hep-th/0610251].

\bibitem{Beisert:2005fw}
  N.~Beisert and M.~Staudacher,
  \textit{``Long-range PSU(2,2|4) Bethe ansaetze for gauge theory and strings,''}
  Nucl.\ Phys.\  B {\bf 727} (2005) 1
  [arXiv:hep-th/0504190].
$\diamondsuit$

\bibitem{Ambjorn:2005wa}
  J.~Ambjorn, R.~A.~Janik and C.~Kristjansen,
  \textit{``Wrapping interactions and a new source of corrections to the spin-chain  /
  string duality,''}
  Nucl.\ Phys.\  B {\bf 736} (2006) 288
  [arXiv:hep-th/0510171].

\bibitem{Arutyunov:2007tc}
  G.~Arutyunov and S.~Frolov,
  \textit{``On String S-matrix, Bound States and TBA,''}
  JHEP {\bf 0712} (2007) 024
  [arXiv:0710.1568 [hep-th]].


\bibitem{Bajnok:2008bm}
  R.~A.~Janik and T.~Lukowski,
  Phys.\ Rev.\  D {\bf 76}, 126008 (2007)
  [arXiv:0708.2208 [hep-th]].
$\diamondsuit$
  M.~P.~Heller, R.~A.~Janik and T.~Lukowski,
  \textit{``A new derivation of Luscher F-term and fluctuations around the giant
  magnon,''}
  JHEP {\bf 0806}, 036 (2008)
  [arXiv:0801.4463 [hep-th]].
$\diamondsuit$
  Z.~Bajnok and R.~A.~Janik,
  \textit{``Four-loop perturbative Konishi from strings and finite size effects for
  multiparticle states,''}
  Nucl.\ Phys.\  B {\bf 807}, 625 (2009)
  [arXiv:0807.0399].

\bibitem{Zamolodchikov:1991et}
  C.~N.~Yang and C.~P.~Yang,
  \textit{``One-dimensional chain of anisotropic spin-spin interactions. I: Proof of
  Bethe's hypothesis for ground state in a finite system,''}
  Phys.\ Rev.\  {\bf 150} (1966) 321.
$\diamondsuit$
  A.~B.~Zamolodchikov,
  \textit{``On the thermodynamic Bethe ansatz equations for reflectionless ADE
  scattering theories,''}
  Phys.\ Lett.\  B {\bf 253}, 391 (1991).
$\diamondsuit$
  N.~Dorey,
\textit{``Magnon bound states and the AdS/CFT correspondence,''}
  J.\ Phys.\ A  {\bf 39}, 13119 (2006)
  [arXiv:hep-th/0604175].
$\diamondsuit$
M.~Takahashi, \textit{``Thermodynamics of one-dimensional solvable models"},
Cambridge University Press, 1999.
$\diamondsuit$
F.H.L. Essler, H.Frahm, F.G\"ohmann, A. Kl\"umper and V. Korepin,
\textit{"The One-Dimensional Hubbard Model"},  Cambridge University Press, 2005.
$\diamondsuit$
%
  V.~V.~Bazhanov, S.~L.~Lukyanov and A.~B.~Zamolodchikov,
  \textit{``Quantum field theories in finite volume: Excited state energies,''}
  Nucl.\ Phys.\  B {\bf 489}, 487 (1997)
  [arXiv:hep-th/9607099].
$\diamondsuit$
  P.~Dorey and R.~Tateo,
  \textit{``Excited states by analytic continuation of TBA equations,''}
  Nucl.\ Phys.\  B {\bf 482}, 639 (1996)
  [arXiv:hep-th/9607167].
$\diamondsuit$
  D.~Fioravanti, A.~Mariottini, E.~Quattrini and F.~Ravanini,
  \textit{``Excited state Destri-De Vega equation for sine-Gordon and restricted
  sine-Gordon models,''}
  Phys.\ Lett.\  B {\bf 390}, 243 (1997)
  [arXiv:hep-th/9608091].
$\diamondsuit$
  A.~G.~Bytsko and J.~Teschner,
\textit{``Quantization of models with non-compact quantum group symmetry: Modular
  XXZ magnet and lattice sinh-Gordon model,''}
  J.\ Phys.\ A  {\bf 39} (2006) 12927
  [arXiv:hep-th/0602093].
$\diamondsuit$
  H.~Saleur and B.~Pozsgay,
\textit{``Scattering and duality in the 2 dimensional $OSP(2|2)$ Gross Neveu and sigma
  models,''}
  arXiv:0910.0637.

\bibitem{Gromov:2008gj}
  N.~Gromov, V.~Kazakov and P.~Vieira,
  ``Finite Volume Spectrum of 2D Field Theories from Hirota Dynamics,''
  arXiv:0812.5091 [hep-th].


\bibitem{Fiamberti:2008sh}
  F.~Fiamberti, A.~Santambrogio, C.~Sieg and D.~Zanon,
  \textit{``Anomalous dimension with wrapping at four loops in N=4 SYM,''}
  Nucl.\ Phys.\  B {\bf 805}, 231 (2008)
  [arXiv:0806.2095].
$\diamondsuit$
V.~N.~Velizhanin,
  \textit{``Leading transcedentality contributions to the four-loop universal anomalous
  dimension in N=4 SYM,''}
  arXiv:0811.0607.



\bibitem{Fiamberti:2009jw}
  F.~Fiamberti, A.~Santambrogio and C.~Sieg,
  \textit{``Five-loop anomalous dimension at critical wrapping order in N=4 SYM,''}
  arXiv:0908.0234 .


\bibitem{Gromov:2009bc}
  N.~Gromov, V.~Kazakov, A.~Kozak and P.~Vieira,
\textit{``Integrability for the Full Spectrum of Planar AdS/CFT II,''}
    arXiv:0902.4458 .

\bibitem{Bombardelli:2009ns}
  D.~Bombardelli, D.~Fioravanti and R.~Tateo,
\textit{``Thermodynamic Bethe Ansatz for planar AdS/CFT: a proposal,''}
  J.\ Phys.\ A  {\bf 42}, 375401 (2009)
  [arXiv:0902.3930].
$\diamondsuit$
  G.~Arutyunov and S.~Frolov,
\textit{``Thermodynamic Bethe Ansatz for the $AdS_5 \times S^5$ Mirror Model,''}
  JHEP {\bf 0905}, 068 (2009)
  [arXiv:0903.0141].






\bibitem{Roiban:2009aa}
  R.~Roiban and A.~A.~Tseytlin,
  \textit{``Quantum strings in $AdS_5 \times S^5$: strong-coupling corrections to dimension of
  Konishi operator,''}
  arXiv:0906.4294.
$\diamondsuit$
  A.~A.~Tseytlin,
 \textit{``Quantum strings in $AdS_5 \times S^5$ and AdS/CFT duality,''}
  arXiv:0907.3238 .

\bibitem{Arutyunov:2009zu}
  G.~Arutyunov and S.~Frolov,
   \textit{``String hypothesis for the $AdS_5 \times S^5$ mirror,''}
  JHEP {\bf 0903} (2009) 152
  [arXiv:0901.1417 [hep-th]].

\bibitem{BF}
  N.~Beisert and L.~Freyhult,
\textit{``Fluctuations and energy shifts in the Bethe ansatz,''}
  Phys.\ Lett.\  B {\bf 622} (2005) 343
  [arXiv:hep-th/0506243].


\bibitem{Gromov:2007ky}
  N.~Gromov and P.~Vieira,
  \textit{``Complete 1-loop test of AdS/CFT,''}
  JHEP {\bf 0804} (2008) 046
  [arXiv:0709.3487].
\bibitem{Gromov:2007cd}
  N.~Gromov and P.~Vieira,
  \textit{``Constructing the AdS/CFT dressing factor,''}
  Nucl.\ Phys.\  B {\bf 790} (2008) 72
  [arXiv:hep-th/0703266].
\bibitem{Gromov:2007aq}
  N.~Gromov and P.~Vieira,
  \textit{``The $AdS_5 \times S^5$ superstring quantum spectrum from the algebraic curve,''}
  Nucl.\ Phys.\  B {\bf 789} (2008) 175
  [arXiv:hep-th/0703191].
\bibitem{Gromov:2008ec}
  N.~Gromov, S.~Schafer-Nameki and P.~Vieira,
  \textit{``Efficient precision quantization in AdS/CFT,''}
  JHEP {\bf 0812} (2008) 013
  [arXiv:0807.4752 ].

\bibitem{Janik:2007wt}
  R.~A.~Janik and T.~Lukowski,
  Phys.\ Rev.\  D {\bf 76}, 126008 (2007)
  [arXiv:0708.2208 [hep-th]].

\bibitem{Gromov:2008ie}
  N.~Gromov, S.~Schafer-Nameki and P.~Vieira,
  \textit{``Quantum Wrapped Giant Magnon,''}
  Phys.\ Rev.\  D {\bf 78} (2008) 026006
  [arXiv:0801.3671].
\bibitem{Gromov:2009zz}
  N.~Gromov,
  \textit{``Integrability in AdS/CFT correspondence: Quasi-classical analysis,''}
  J.\ Phys.\ A  {\bf 42} (2009) 254004.
\bibitem{SchaferNameki:2006gk}
  S.~Schafer-Nameki,
  \textit{``Exact expressions for quantum corrections to spinning strings,''}
  Phys.\ Lett.\  B {\bf 639} (2006) 571
  [arXiv:hep-th/0602214].
$\diamondsuit$
  S.~Schafer-Nameki, M.~Zamaklar and K.~Zarembo,
  \textit{``How accurate is the quantum string Bethe ansatz?,''}
  JHEP {\bf 0612} (2006) 020
  [arXiv:hep-th/0610250].

\bibitem{Sutherland:1995zz}
  B.~Sutherland,
\textit{``Low-Lying Eigenstates of the One-Dimensional Heisenberg Ferromagnet for any
  Magnetization and Momentum,''}
  Phys.\ Rev.\ Lett.\  {\bf 74} (1995) 816.

\bibitem{Beisert:2003xu}
  N.~Beisert, J.~A.~Minahan, M.~Staudacher and K.~Zarembo,
\textit{``Stringing spins and spinning strings,''}
  JHEP {\bf 0309} (2003) 010
  [arXiv:hep-th/0306139].

\bibitem{Tsuboi:1997iq}
Z.~Tsuboi,
\textit{``Analytic Bethe ansatz and functional equations for Lie superalgebra
$sl(r+1|s+1)$,''}
J.\ Phys.\ A {\bf 30}, 7975 (1997);
$\diamondsuit$
  N.~Beisert,
\textit{``The Analytic Bethe Ansatz for a Chain with Centrally Extended $su(2|2)$ Symmetry,''}
  J.\ Stat.\ Mech.\  {\bf 0701} (2007) P017
  [arXiv:nlin/0610017].


\bibitem{Kazakov:2007fy}
  V.~Kazakov, A.~S.~Sorin and A.~Zabrodin,
  \textit{``Supersymmetric Bethe ansatz and Baxter equations from discrete Hirota
  dynamics,''}
  Nucl.\ Phys.\  B {\bf 790}, 345 (2008)
  [arXiv:hep-th/0703147].


\bibitem{Arutyunov:2009ce}
  G.~Arutyunov, M.~de Leeuw and A.~Torrielli,
\textit{``Universal blocks of the AdS/CFT Scattering Matrix,''}
  JHEP {\bf 0905} (2009) 086
  [arXiv:0903.1833].

\bibitem{FS4}
  N.~Beisert, A.~A.~Tseytlin and K.~Zarembo,
   \textit{``Matching quantum strings to quantum spins: One-loop vs. finite-size corrections,''}
  Nucl.\ Phys.\  B {\bf 715}, 190 (2005)
  [arXiv:hep-th/0502173].
$\diamondsuit$
  R.~Hernandez, E.~Lopez, A.~Perianez and G.~Sierra,
   \textit{``Finite size effects in ferromagnetic spin chains and quantum  corrections
  to classical strings,''}
  JHEP {\bf 0506}, 011 (2005)
  [arXiv:hep-th/0502188].
$\diamondsuit$
  S.~Schafer-Nameki, M.~Zamaklar and K.~Zarembo,
   \textit{``Quantum corrections to spinning strings in $AdS_5 \times S^5$ and Bethe  ansatz:
 A comparative study,''}
  JHEP {\bf 0509}, 051 (2005)
  [arXiv:hep-th/0507189].
$\diamondsuit$
  N.~Gromov and V.~Kazakov,
\textit{``Double scaling and finite size corrections in sl(2) spin chain,''}
  Nucl.\ Phys.\  B {\bf 736}, 199 (2006)
  [arXiv:hep-th/0510194].

\bibitem{FS1}
  N.~Beisert and A.~A.~Tseytlin,
  \textit{``On quantum corrections to spinning strings and Bethe equations,''}
  Phys.\ Lett.\  B {\bf 629}, 102 (2005)
  [arXiv:hep-th/0509084].


\bibitem{Hernandez:2006tk}
R.~Hernandez and E.~Lopez,
{\it {Quantum corrections to the string Bethe
  ansatz}},
 {\em JHEP} {\bf 07} (2006) 004
$\diamondsuit$
  L.~Freyhult and C.~Kristjansen,
  {\it``A universality test of the quantum string Bethe ansatz,''}
  Phys.\ Lett.\  B {\bf 638}, 258 (2006)
  [arXiv:hep-th/0604069].



\bibitem{Frolov:2002av}
  S.~Frolov and A.~A.~Tseytlin,
\textit{``Semiclassical quantization of rotating superstring in $AdS_5 \times S^5$,''}
  JHEP {\bf 0206}, 007 (2002)
  [arXiv:hep-th/0204226].

\bibitem{Beisert:2005di}
  N.~Beisert, V.~A.~Kazakov, K.~Sakai and K.~Zarembo,
  \textit{``Complete spectrum of long operators in N = 4 SYM at one loop,''}
  JHEP {\bf 0507}, 030 (2005)
  [arXiv:hep-th/0503200].


\bibitem{C1}
  S.~Frolov and A.~A.~Tseytlin,
\textit{``Multi-spin string solutions in $AdS_5 \times S^5$,''}
  Nucl.\ Phys.\  B {\bf 668}, 77 (2003)
  [arXiv:hep-th/0304255].
$\diamondsuit$
  G.~Arutyunov, J.~Russo and A.~A.~Tseytlin,
\textit{``Spinning strings in $AdS_5 \times S^5$: New integrable system relations,''}
  Phys.\ Rev.\  D {\bf 69}, 086009 (2004)
  [arXiv:hep-th/0311004].

\bibitem{C2}
  S.~Frolov and A.~A.~Tseytlin,
\textit{``Quantizing three-spin string solution in $AdS_5 \times S^5$,''}
  JHEP {\bf 0307}, 016 (2003)
  [arXiv:hep-th/0306130].
$\diamondsuit$
  S.~A.~Frolov, I.~Y.~Park and A.~A.~Tseytlin,
  Phys.\ Rev.\  D {\bf 71} (2005) 026006
  [arXiv:hep-th/0408187].
$\diamondsuit$
  I.~Y.~Park, A.~Tirziu and A.~A.~Tseytlin,
\textit{``Spinning strings in $AdS_5 \times S^5$: One-loop correction to energy in  SL(2) sector,''}
  JHEP {\bf 0503}, 013 (2005)
  [arXiv:hep-th/0501203].



\bibitem{Metsaev:1998it}
  R.~R.~Metsaev and A.~A.~Tseytlin,
  ``Type IIB superstring action in AdS(5) x S(5) background,''
  Nucl.\ Phys.\  B {\bf 533}, 109 (1998)
  [arXiv:hep-th/9805028].

\bibitem{Kazakov:2004nh}
  V.~A.~Kazakov and K.~Zarembo,
\textit{``Classical / quantum integrability in non-compact sector of AdS/CFT,''}
  JHEP {\bf 0410} (2004) 060
  [arXiv:hep-th/0410105].


\bibitem{Vicedo:2008jy}
  B.~Vicedo,
  \textit{``Semiclassical Quantisation of Finite-Gap Strings,''}
  JHEP {\bf 0806}, 086 (2008)
  [arXiv:0803.1605].

\bibitem{Bajnok:2008qj}
  Z.~Bajnok, R.~A.~Janik and T.~Lukowski,
  \textit{``Four loop twist two, BFKL, wrapping and strings,''}
  Nucl.\ Phys.\  B {\bf 816} (2009) 376
  [arXiv:0811.4448].


\bibitem{Bajnok:2009vm}
  Z.~Bajnok, A.~Hegedus, R.~A.~Janik and T.~Lukowski,
\textit{``Five loop Konishi from AdS/CFT,''}
  arXiv:0906.4062 .


\bibitem{Tsuboi:2009ud}
  Z.~Tsuboi,
\textit{``Solutions of the T-system and Baxter equations for supersymmetric spin
  chains,''}
  arXiv:0906.2039.
$\diamondsuit$
  A.~Hegedus,
\textit{``Discrete Hirota dynamics for AdS/CFT,''}
  arXiv:0906.2546 .

\bibitem{ABJM}
  O.~Aharony, O.~Bergman, D.~L.~Jafferis and J.~Maldacena,
  \textit{``N=6 superconformal Chern-Simons-matter theories, M2-branes and their
  gravity duals,''}
  JHEP {\bf 0810}, 091 (2008)
  [arXiv:0806.1218].

\bibitem{Minahan:2009aq}
  J.~A.~Minahan, O.~O.~Sax and C.~Sieg,
  \textit{``Magnon dispersion to four loops in the ABJM and ABJ models,''}
  arXiv:0908.2463.

\bibitem{Arutyunov:2009kf}
  G.~Arutyunov and S.~Frolov,
  \textit{``The Dressing Factor and Crossing Equations,''}
  arXiv:0904.4575.


\end{thebibliography}
\end{document}